\documentclass[reqno,11pt]{amsart}\usepackage[]{graphicx}\usepackage[]{color}
\makeatletter
\def\maxwidth{ %
  \ifdim\Gin@nat@width>\linewidth
    \linewidth
  \else
    \Gin@nat@width
  \fi
}
\makeatother

\definecolor{fgcolor}{rgb}{0.345, 0.345, 0.345}

\usepackage{framed}
\makeatletter
 {\par\unskip\endMakeFramed%
 \at@end@of@kframe}
\makeatother

\definecolor{shadecolor}{rgb}{.97, .97, .97}
\definecolor{messagecolor}{rgb}{0, 0, 0}
\definecolor{warningcolor}{rgb}{1, 0, 1}
\definecolor{errorcolor}{rgb}{1, 0, 0}
\usepackage{alltt}

\usepackage{accents,setspace,graphicx,srcltx,enumitem}
\usepackage[round]{natbib}
\usepackage{subfigure}
\usepackage{bbm}
\usepackage{multirow}
\usepackage{amsmath}
\usepackage{amssymb}
\usepackage{amsfonts}
\usepackage[english]{babel}
\allowdisplaybreaks

\usepackage{MnSymbol}  
\usepackage{hyperref}
\usepackage{adjustbox}
\usepackage{array}

\newtheorem{theorem}{Theorem}[section]
\newtheorem{proposition}{Proposition}[section]
\newtheorem{assumption}{Assumption}[section]

  \theoremstyle{remark}

\DeclareMathOperator*{\argmax}{arg\,max}

      \newcommand{\Tau}{\mathrm{T}}
      \newcommand{\field}[1]{\mathbb{#1}}
      \newcommand{\R}{\field{R}}

      \newcommand{\tsigma}{\tilde{\sigma}}
      \newcommand{\Z}{\mathbf{0}}

\title{Maximum Likelihood Estimation of Stochastic Frontier Models with Endogeneity}

\date{\today}
\author{Samuele Centorrino}
\address[S.\ Centorrino, Corresponding author]{Economics Department, State University of New York at Stony Brook, USA.}
\email[S.\ Centorrino]{\texttt{\textup{samuele.centorrino@stonybrook.edu}}} 
\author{Mar\'ia P\'erez-Urdiales}
\address[M.\ P\'erez-Urdiales]{Economics Department, State University of New York at Stony Brook, USA.}
\email{\texttt{\textup{maria.perez-urdiales@stonybrook.edu}}}

\thanks{We wish to thank the Editor, Elie Tamer, the Associate Editor, two anonymous referees, Christopher F. Parmeter, and participants to the North American Productivity Workshop XI for their comments and remarks.}

\IfFileExists{upquote.sty}{\usepackage{upquote}}{}
\begin{document}

\begin{abstract}
We propose and study a maximum likelihood estimator of stochastic frontier models with endogeneity in cross-section data when the composite error term may be correlated with inputs and environmental variables. Our framework is a generalization of the normal half-normal stochastic frontier model with endogeneity. We derive the likelihood function in closed form using three fundamental assumptions: the existence of control functions that fully capture the dependence between regressors and unobservables; the conditional independence of the two error components given the control functions; and the conditional distribution of the stochastic inefficiency term given the control functions being a folded normal distribution. We also provide a Battese-Coelli estimator of technical efficiency. Our estimator is computationally fast and easy to implement. We study some of its asymptotic properties, and we showcase its finite sample behavior in Monte-Carlo simulations and an empirical application to farmers in Nepal.\\



\end{abstract}

\maketitle

\onehalfspacing

\section{Introduction} \label{sec:intro}

Endogeneity in the stochastic frontier framework has received increasing attention in recent work (\citealp[see][among others]{kutlu2010,tran2013,tran2015,karakaplan2017,amsler2016,Lai2018}, and \citealp{kumbhakar2020a,kumbhakar2020b}, for a review). Most contributions focus on the correlation between the regressors and the two-sided error component while ignoring the potential dependence between regressors and the stochastic inefficiency component. However, if producers have some information about their inefficiency level, they can use it to guide their choice of inputs and environmental variables (e.g., managerial characteristics). That is, there may be factors, observable to the firm but unobservable to the econometrician, which affect both the choice of regressors, and the level of inefficiency \citep[see][for a similar argument]{cazals2016}. 

In this paper, we consider a stochastic frontier model in which both the two-sided error and the stochastic inefficiency terms are allowed to be correlated with inputs and environmental variables. These endogenous variables are restricted to be continuous. The production frontier can be linear or nonlinear, and the inefficiency term satisfies the scaling property. That is, it can be decomposed into a stochastic efficiency term and a scaling function that depends on environmental variables \citep{alvarez2006}. We achieve identification by allowing for a vector of control functions that fully captures the dependence between the composite error term and the endogenous variables.

To the best of our knowledge, models that explicitly allow for dependence between the stochastic inefficiency term, inputs, and environmental variables have only been studied by \citet{amsler2017}.\footnote{\citet{karakaplan2017} do not directly consider the potential endogeneity of the inefficiency term. They instead model the potential dependence between the two-sided error term and the inefficiency term through observables, while we model such dependence through unobservables.} In their paper, the marginal distribution of the statistical noise is taken to be a normal distribution, and the marginal distribution of the stochastic inefficiency term to be a half-normal distribution. The two stochastic terms are potentially correlated. The dependence between observables and unobservables is modeled using copula functions, which are cleverly constructed from the marginal distributions of the unobservables. However, the likelihood function cannot be written in closed form, and the authors need to resort to simulations to obtain an estimator of the model's parameters. This approach prevents a clear analysis of identification, estimation, and inference. Moreover, simulated methods can be biased and have a higher variance in finite samples, especially when the number of simulations is not chosen appropriately with the sample size \citep{gourieroux1996}. Finally, when both inputs and environmental variables are potentially correlated with the inefficiency term, they cannot obtain an estimator of technical efficiency.

Our framework seeks to avoid these potential pitfalls. In particular, we are able to obtain the maximum likelihood function in closed form. This allows us to further the analysis of identification of this model and propose a simple and computationally fast estimation of the model's parameters. We also offer a generalization of the \citet{battese1988} estimator of technical efficiency.

While our main statistical model is similar to the one in \citet{amsler2017}, two fundamental assumptions deviate from their framework. First, we assume that the two-sided error term and the inefficiency term are independent conditional on a vector of control functions. This first assumption allows us to write the conditional density of the composite error as the convolution of the conditional densities of the statistical noise and the inefficiency term, respectively. Second, we assume that the conditional distribution of the baseline stochastic inefficiency term given the control functions is a folded normal distribution \citep{leone1961,sundberg1974}. The latter assumption is convenient for two main reasons. On the one hand, it allows us to capture the dependence between inefficiency and endogenous variables through a vector of what we refer to as dependence parameters, $\rho_{U}$, which take values in the hypercube $[-1,1]$. These parameters measure the fraction of inefficiency observed by the producer but not by the econometrician, which may influence the choice of inputs and environmental factors and is confounded with the observed level of the regressors. On the other hand, the conditional normal-folded normal model provides a natural generalization of the normal half-normal model to the case when inputs and environmental variables are endogenous. This is because the folded normal pdf collapses to the density of a half-normal random variable when $\rho_{U} = \Z$, where $\Z$ is a vector of zeros. That is, when stochastic inefficiency is unobservable to producers and hence cannot influence their decision.

Our analysis of identification and estimation focuses on the dependence parameters $\rho_U$. Because of the properties of the folded normal distribution, only the magnitude of the components of $\rho_U$ is identified. However, their sign cannot be identified \citep{sundberg1974,schmidt1980}.\footnote{The folded normal distribution can be thought of as a normal distribution ``folded" at zero by taking the absolute value. Suppose we take a mean-zero normal random variable $\eta$, and then generate two standard normal random variables $U_1$ and $U_2$, which have correlation $-0.5$, and $0.5$ with $\eta$, respectively. When we ``fold" both $U_1$ and $U_2$ by taking their absolute values, we have that $\vert U_1 \vert$ has the same conditional distribution of $\vert U_2 \vert$. Identification of the sign of $\rho_U$ is thus not feasible.} Hence, the likelihood function has two isolated maxima, which are symmetric about a local extremum at zero. We deal with this identification issue by imposing a sign normalization which amounts to restricting one of the components of $\rho_U$ to lay in the positive orthant. When $\rho_U = \Z$, the likelihood function has a unique extremum. However, when $\rho_U = \Z$ the score is identically equal to zero, and our model is not first-order identified. We are nonetheless able to show that our model is second-order identified.

Moreover, since one of the components of $\rho_U$ lays at the boundary of the parameters' space and the Hessian matrix is singular when $\rho_U = \Z$, our estimator has a non-standard asymptotic distribution, and its rate of convergence is slower than $\sqrt{n}$, where $n$ is the sample size. We provide the asymptotic distribution of our estimator in all these cases following the framework of \citet{andrews1999} and \citet{rot2000}. Finally, we briefly discuss potential ways to conduct inference on $\rho_U$.

Throughout the paper, we assume that parameters other than $\rho_U$ are first-order locally identified, and thus $\sqrt{n}$-estimable \citep{sargan1983}. This assumption implies, in particular, that the variance of the inefficiency term is strictly positive. We defer to future research the study of this model when such an assumption fails \citep[see, e.g.][for identifying and estimating the classical stochastic frontier model with lack of first-order identification]{lee1993}.

The paper is structured as follows. In Section \ref{sec:model}, we discuss the statistical model and provide the main steps for the construction of the likelihood function. We further consider identification, estimation and inference. In Section \ref{sec:montecarlo}, we provide simulation evidence of the finite sample properties of our estimator. We show that our estimator performs better than the copula method of \citet{amsler2017} especially for estimating the variance of the stochastic inefficiency term. In Section \ref{sec:empiricalapp}, we apply our methodology to the agricultural sector in Nepal. We show that accounting for endogeneity substantially changes the conclusions of the empirical analysis. In particular, our estimator detects considerable variation in the efficiency scores which is not found when regressors are taken to be exogenous.

\section{Statistical Model} \label{sec:model}

We study a general version of the model usually considered in this literature. The logarithm of the output, $Y$, is determined by some known function, $m(\cdot,\cdot)$, which depends on a vector of $p \geq 1$ inputs, $X$, and parameters, $\beta$; and by a composite error term $\varepsilon = V - U$, where $V$ represents a stochastic component; and $U \geq 0$ is the so-called inefficiency term. We thus have
\begin{equation}\label{stmod}
Y=m(X,\beta)+V - U,
\end{equation}
where $U$ captures the producer's shortfall from the production frontier. 

Additionally, we fix $U = U_0g(Z,\delta)$, where $U_0\geq 0$ is a stochastic inefficiency component and $g(\cdot,\cdot)$ is a known strictly positive scaling function, which depends on some additional \textit{environmental variables} $Z \in \R^k$, with $k \geq 0$, through parameters $\delta$ \citep{simar1994,alvarez2006}. The scaling function further satisfies the normalization condition $g(0,\delta) = 1$. $X$ and $Z$ may have some common elements, but they must have at least one non-overlapping component.

Thus, we finally have
\begin{equation}\label{stmod2}
Y=m(X,\beta)+V - U_0g(Z,\delta).
\end{equation}

A maximum likelihood estimator of $(\beta,\delta)$ is based on the assumption that the composite error component $(V,U_0)$ is independent of $(X,Z)$, with $(U_0,V)$ mutually independent; $V$ following a normal distribution with a constant variance, and $U_0$ following a normal distribution truncated at $0$ \citep[so-called positive half-normal distribution, see][]{aigner1977,schmidt1979,schmidt1980,horrace2005}. While a consistent estimation of $(\beta,\delta)$ can also be obtained without these strong distributional assumptions \citep{simar1994,tran2013}, these assumptions are necessary to learn something about the variance of the inefficiency term, $U_0$. We are often interested in estimating each producer's distance from the frontier \citep{battese1988}. This can be easily done when the marginal distributions of $V$ and $U_0$ are taken to be known. 

The literature has long recognized that inputs may be simultaneously chosen with the output, and thus potentially correlated with the composite error term \citep[see][for a full description of the statistical issues in this context]{mundlak1961,schmidt1984}. Similarly, the producer may decide environmental variables depending on characteristics that are observable to her but not to the econometrician.

To deal with endogenous variables, we need a vector of instruments that are correlated with the endogenous components but independent of the composite error term \citep[see][for the impact of several exogeneity assumptions on identification in SFA]{amsler2016}. To simplify our presentation, we take all variables in $(X,Z)$ to be endogenous. The extension to the case when we have some endogenous and some exogenous components can be handled similarly. 

We consider the following auxiliary regression models
\begin{align*}
X =& W\gamma_X + \eta_X\\
Z =& W\gamma_Z + \eta_Z,
\end{align*}
where $\eta = (\eta^\prime_X,\eta^\prime_Z)^\prime \in \R^{p+k}$ is a random vector of error components, and $W\in \R^q$ is a vector of instrumental variables, with $q \geq p + k$.

Our approach is based on a control function assumption. That is, we assume that all the dependence between $(X,Z)$ and $(V,U_0)$ is captured by $\eta$ \citep{newey1999,newey2009,wooldridge2009}. Moreover, we assume that the instruments are strongly exogenous, that is, independent of the composite error term. Given a triplet of random variables $U_0$, $V$ and $\eta$, we use the notation $U_0 \upmodels V$ to indicate that $U_0$ is independent of $V$; and the notation $U_0 \upmodels V \vert \eta$ to indicate that $U_0$ is independent of $V$ conditional on $\eta$. 

Our independence assumptions can be formally stated as follows:

\begin{assumption}~\label{ass:cindepeta}
$W \upmodels (V,U_0,\eta)$,
\end{assumption}

\begin{assumption}~\label{ass:cindeuv}
$U_0 \upmodels V \vert \eta$. 
\end{assumption}

Assumption \ref{ass:cindepeta} implies strong exogeneity of the instruments; and that the control function, $\eta$, captures all the dependence between $(X,Z)$ and $(U_0,V)$. That is, $(X,Z) \upmodels (U_0,V) \vert \eta$.

Assumption \ref{ass:cindeuv} implies that, if any dependence exists between $V$ and $U_0$, it has to happen through the vector $\eta$. This assumption reduces to the standard assumption of $U_0 \upmodels V$ when both $X$ and $Z$ are taken to be exogenous \citep[Sec. 3.2, p. 64]{kumbhakar2003}.

Assumptions \ref{ass:cindepeta} and \ref{ass:cindeuv} imply that 
\[
f_{V,U_0,\eta \vert W} (v,u,\eta \vert W) = f_{V,U_0,\eta} (v,u,\eta )= f_{V,\eta} (v ,\eta ) f_{U_0\vert \eta} (u \vert\eta ),
\]
where $f$ denotes a probability density function. To construct a maximum likelihood estimator (MLE), we let $\eta \sim N(0,\Sigma_\eta)$, where $\Sigma_\eta$ is a symmetric, positive definite covariance matrix. We also let $D_{\eta}$ be a $p+ k \times p + k$ diagonal matrix whose diagonal entries are the standard deviations of $\eta$. We can write that $\Sigma_\eta = D_{\eta}C_\eta D_{\eta}$, where $C_\eta$ is the symmetric, positive definite correlation matrix of the vector $\eta$. That is, a matrix with diagonal equal to $1$ and the other elements in the interval $(-1,1)$. 

An additional requirement for the construction of a full information MLE is that
\[
\begin{pmatrix} V \\ \eta \end{pmatrix}\sim N\left(\begin{bmatrix} 0\\ 0 \end{bmatrix}, \begin{bmatrix} \sigma^2_V &\sigma_V  \rho^\prime_V D_\eta  \\ D_{\eta} \rho_V \sigma_V &  \Sigma_{\eta}\end{bmatrix} \right),
\]
where $\rho_{V}$ is a vector of correlation coefficients between $V$ and all components of $\eta$, and $\sigma^2_V$ is the variance of $V$ \citep[see also][]{kutlu2010}. 

The main difficulty lies in the specification of the joint density of $(U_0,\eta)$ such that its marginal distributions are a half-normal and a joint normal, respectively, and the dependence between the two is captured by only one parameter. If one specifies a joint normal distribution for the random vector $(U^\ast_0,\eta)$ and then takes $U_0 = \vert U^\ast_0\vert$, the marginal distributions of $U_0$ and $\eta$ are the correct marginal distributions. This construction also creates dependence between $U_0$ and $\eta$. Here, we argue that the conditional distribution of $U_0$ given $\eta$ can be written in such a way that this dependence is captured by only one vector of parameters which, we refer to as dependence parameters, and we denote as $\rho_U$.
Let
\[
\begin{pmatrix} U^\ast_0 \\ \eta \end{pmatrix}\sim N\left(\begin{bmatrix} 0\\ 0 \end{bmatrix}, \begin{bmatrix} \sigma^2_U & \sigma_U\rho^\prime_{U} D_\eta\\ D_\eta\rho_{U} \sigma_U &  \Sigma_{\eta}\end{bmatrix} \right),
\]
where $\rho_{U}$ is a vector of \textit{correlations} between $U^\ast_0$ and $\eta$, and $\sigma^2_U$ is the variance of $U^\ast_0$. The conditional density of $U^\ast_0$ given $\eta$ is
\[
f_{U^\ast_0 \vert \eta } (u \vert\eta ) = \frac{1}{\sqrt{2\pi \sigma_U^2 (1 - \rho^\prime_{U}C^{-1}_\eta \rho_{U})}} \exp \left( -\frac{(u - \sigma_U \rho^\prime_{U} C^{-1}_\eta D^{-1}_\eta \eta)^2}{2 \sigma_U^2 (1 - \rho^\prime_{U}C^{-1}_\eta \rho_{U})}\right).
\]
Thus, we have that 
\[
P\left( U_0 \leq u \vert \eta \right) = P\left( U^\ast_0 \leq u \vert \eta \right) - P\left( U^\ast_0 \leq -u \vert \eta \right).
\]
Taking the derivative of the last equality with respect to $u$ on both sides, we obtain that the density of $U_0$ given $\eta$ is equal to 
\[
f_{U_0 \vert \eta }(u \vert \eta )= f_{U^\ast_0 \vert \eta } (u \vert\eta ) + f_{U^\ast_0 \vert \eta } (-u \vert\eta ).
\]
Therefore, the conditional density function of $U_0$ is
\begin{equation} \label{eq:densuclosed}
f_{U_0 \vert \eta}(u \vert \eta) = \frac{1}{\sqrt{2\pi \sigma_U^2 (1 - \rho^\prime_{U}C^{-1}_\eta \rho_{U})}}\left\lbrace \exp \left( -\frac{(u -  \sigma_U \rho^\prime_{U} C^{-1}_\eta D^{-1}_\eta \eta)^2}{2\sigma_U^2 (1 - \rho^\prime_{U}C^{-1}_\eta \rho_{U})}\right) + \exp \left( -\frac{(u +  \sigma_U \rho^\prime_{U} C^{-1}_\eta D^{-1}_\eta \eta)^2}{2\sigma_U^2 (1 - \rho^\prime_{U}C^{-1}_\eta \rho_{U})}\right)\right\rbrace,
\end{equation}
which is the pdf of a folded normal distribution \citep{leone1961}. When we impose that $\rho_{U}$ is a vector of zeros, that is, when there is no dependence between the regressors and the inefficiency term, the conditional distribution in \eqref{eq:densuclosed} reduces to
\[
f_{U_0}(u) =  \frac{2}{\sqrt{2\pi\sigma^2_U}} \exp \left( -\frac{u^2}{2\sigma_U^2}\right),
\]
which is the density of a half-normal distribution. We also show in Appendix \ref{sec:appA} that the marginal density of $U_0$ obtained from this construction is a half-normal density.

\begin{figure}[!h]
    \begin{center}
      \includegraphics[scale=.5,trim={1in 2in 1in 1.5in}]{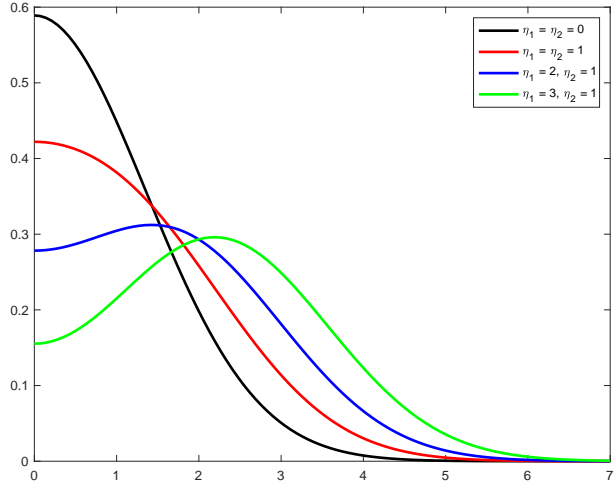}
      \caption{\label{fig:cpdfu}Conditional density of $U_0$ given $\eta$.}
    \end{center}
  \end{figure}

Figure \ref{fig:cpdfu} depicts the conditional folded normal pdf when $\eta$ is a bivariate random vector with unit variance and correlation coefficient equal to $0.5$, $\sigma^2_U = 2.752$, and $\rho_U = (0.5,0.5)^\prime$. For fixed parameters, the pdf is symmetric in $\eta$, in the sense that the shape of the density for $\eta =e$ is the same as for $\eta = -e$, for any real-valued vector $e$.

For a given $\eta$, this implies that the density is invariant to changes in sign of the vector of dependence parameters $\rho_U$. That is, the conditional density of $U_0$ generated under a certain dependence vector $\rho_U$ is equal to the conditional density of $U_0$ when the dependence vector is $-\rho_U$. This is a well-known equivalence property of the folded normal distribution \citep[see][among others]{sundberg1974}.

The construction of the likelihood function is thus based on the following
\begin{assumption}~\label{ass:distribution}
\begin{itemize}
\item[(i)] $\eta \sim N\left( 0, \Sigma_\eta \right)$, where $\Sigma_\eta = D_\eta C_\eta D_\eta$ is a positive definite and symmetric covariance matrix; 
\item[(ii)] $V \vert \eta \sim N\left(\sigma_V \rho^\prime_V C^{-1}_\eta D^{-1}_\eta \eta, \sigma^2_V (1 - \rho^\prime_V C^{-1}_\eta \rho_V)  \right)$.
\item[(iii)] $U_0 \vert \eta \sim  FN\left(\sigma_U \rho^\prime_U C^{-1}_\eta D^{-1}_\eta \eta, \sigma^2_U (1 - \rho^\prime_U C^{-1}_\eta \rho_U) \right)$, where $FN$ denotes a folded-normal distribution with location parameter  $\sigma_U \rho^\prime_U C^{-1}_\eta D^{-1}_\eta \eta$; and scale parameter $\sigma^2_U (1 - \rho^\prime_U C^{-1}_\eta \rho_U)$. 
\end{itemize}
\end{assumption}
Finally, because of Assumption \ref{ass:cindepeta} and the strict positivity of the function $g(\cdot,\cdot)$, the conditional distribution of $U = U_0 g(Z,\delta)$ given $\eta$ can be written as 
\begin{equation} \label{eq:scaleequ}
P(U \leq u \vert \eta) = P(U_0 \leq \left( g(Z,\delta )\right)^{-1}u \vert \eta ),
\end{equation}
and it is therefore a simple scaled version of the distribution of $U_0$ given $\eta$, as in the exogenous case.

We follow the literature on stochastic frontier and define a new random variable $\varepsilon = V - U$ such that
\[
f_{U,\varepsilon\vert \eta} (u,\varepsilon \vert \eta) = f_{V\vert\eta} (\varepsilon + u \vert \eta ) \left( g(Z,\delta )\right)^{-1} f_{U_0\vert \eta }( \left( g(Z,\delta )\right)^{-1} u \vert \eta).
\]
We can thus write
\begin{align*}
f_{V\vert\eta} & (\varepsilon + u \vert \eta ) \left( g(Z,\delta )\right)^{-1} f_{U_0\vert \eta }( \left( g(Z,\delta )\right)^{-1} u  \vert \eta)  \\
=& \frac{1}{2\pi \tilde{\sigma}_U (Z) \tilde{\sigma}_V} \left\lbrace \exp \left( -\frac{(u - \sigma_U g \left( Z,\delta\right) \rho_{U}^\prime C^{-1}_\eta D^{-1}_\eta \eta)^2}{2\tilde{\sigma}^2_U(Z)}- \frac{(\varepsilon + u - \sigma_V \rho_{V}^\prime C^{-1}_\eta D^{-1}_\eta\eta)^2}{2\tilde{\sigma}^2_V}\right) \right. \\
&\quad \left. + \exp \left( -\frac{(u + \sigma_U g \left( Z,\delta\right) \rho_{U}^\prime C^{-1}_\eta D^{-1}_\eta\eta)^2}{2\tilde{\sigma}^2_U(Z)}- \frac{(\varepsilon + u - \sigma_V \rho_{V}^\prime C^{-1}_\eta D^{-1}_\eta \eta)^2}{2\tilde{\sigma}^2_V}\right) \right\rbrace,
\end{align*}
where $\tilde{\sigma}^2_U(Z) = \sigma^2_U g^2\left( Z,\delta\right) \left( 1  -  \rho_{U}^\prime C^{-1}_\eta \rho_{U}\right)$, and $\tilde{\sigma}^2_V = \sigma^2_V \left( 1 - \rho_{V}^\prime C^{-1}_\eta \rho_{V} \right)$. 

By tedious computations that we detail in Appendix \ref{sec:appA}, and after integrating with respect to $U$, we obtain
\begin{align}\label{eq:conddens}
f_{\varepsilon \vert \eta} (\varepsilon \vert \eta) =& \int f_{V\vert\eta} (\varepsilon + u \vert \eta ) \left( g(Z,\delta )\right)^{-1} f_{U_0\vert \eta } (\left( g(Z,\delta )\right)^{-1} u \vert \eta  ) du \nonumber \\
=& \frac{1}{\sqrt{2\pi} \sigma(Z)} \left\lbrace \Phi \left( \frac{g\left( Z,\delta\right) \sigma_U \rho_{U}^\prime C^{-1}_\eta D^{-1}_\eta \eta}{\lambda(Z) \sigma(Z)} -\frac{\lambda(Z) ( \varepsilon - \sigma_V \rho_{V}^\prime C^{-1}_\eta D^{-1}_\eta \eta)}{\sigma(Z)}\right) \times \right. \nonumber\\
& \quad \exp \left( -\frac{(\varepsilon -  \sigma_V \rho_{V}^\prime C^{-1}_\eta D^{-1}_\eta \eta + g\left( Z,\delta\right)  \sigma_U \rho_{U}^\prime C^{-1}_\eta D^{-1}_\eta \eta)^2}{2 \sigma^2(Z)}\right)\nonumber\\
& \quad + \Phi \left( - \frac{g \left( Z,\delta\right)  \sigma_U \rho_{U}^\prime C^{-1}_\eta D^{-1}_\eta \eta}{\lambda(Z)\sigma(Z)} -\frac{\lambda(Z) ( \varepsilon - \sigma_V \rho_{V}^\prime C^{-1}_\eta D^{-1}_\eta \eta)}{\sigma(Z)}\right) \times \nonumber\\
& \quad \left. \exp \left( - \frac{(\varepsilon -\sigma_V \rho_{V}^\prime C^{-1}_\eta D^{-1}_\eta \eta - g \left( Z,\delta\right)  \sigma_U \rho_{U}^\prime C^{-1}_\eta D^{-1}_\eta \eta)^2}{2 \sigma^2(Z)}\right) \right\rbrace\nonumber\\
=& \Phi \left( \frac{ \tsigma^2_V g\left( Z,\delta\right) \sigma_U \rho_{U}^\prime C^{-1}_\eta D^{-1}_\eta \eta}{\tilde{\sigma}_U(Z)  \sigma^2(Z)}\right) \left[ \frac{1}{\sqrt{2\pi} \sigma(Z)} \frac{\Phi \left( \frac{g\left( Z,\delta\right) \sigma_U \rho_{U}^\prime C^{-1}_\eta D^{-1}_\eta \eta}{\lambda(Z) \sigma(Z)} -\frac{\lambda(Z) ( \varepsilon - \sigma_V \rho_{V}^\prime C^{-1}_\eta D^{-1}_\eta \eta)}{\sigma(Z)}\right)}{\Phi \left( \frac{ \tsigma^2_V g\left( Z,\delta\right) \sigma_U \rho_{U}^\prime C^{-1}_\eta D^{-1}_\eta \eta}{\tilde{\sigma}_U(Z)\sigma^2(Z)} \right)} \times \right. \nonumber\\
& \quad \left. \exp \left( -\frac{(\varepsilon -  \sigma_V \rho_{V}^\prime C^{-1}_\eta D^{-1}_\eta \eta + g\left( Z,\delta\right)  \sigma_U \rho_{U}^\prime C^{-1}_\eta D^{-1}_\eta \eta)^2}{2 \sigma^2(Z)}\right) \right] \nonumber\\
& \quad + \Phi \left( -\frac{ \tsigma^2_V g\left( Z,\delta\right) \sigma_U \rho_{U}^\prime C^{-1}_\eta D^{-1}_\eta \eta}{\tilde{\sigma}_U(Z) \sigma^2(Z)} \right) \left[ \frac{1}{\sqrt{2\pi} \sigma(Z)} \frac{\Phi \left( - \frac{g \left( Z,\delta\right)  \sigma_U \rho_{U}^\prime C^{-1}_\eta D^{-1}_\eta \eta}{\lambda(Z)\sigma(Z)} -\frac{\lambda(Z) ( \varepsilon - \sigma_V \rho_{V}^\prime C^{-1}_\eta D^{-1}_\eta \eta)}{\sigma(Z)}\right)}{\Phi \left( -\frac{ \tsigma^2_V g\left( Z,\delta\right) \sigma_U \rho_{U}^\prime C^{-1}_\eta D^{-1}_\eta \eta}{\tilde{\sigma}_U(Z)\sigma^2(Z)}\right)} \times \right. \nonumber\\
& \quad \left. \exp \left( - \frac{(\varepsilon -\sigma_V \rho_{V}^\prime C^{-1}_\eta D^{-1}_\eta \eta - g \left( Z,\delta\right)  \sigma_U \rho_{U}^\prime C^{-1}_\eta D^{-1}_\eta \eta)^2}{2 \sigma^2(Z)}\right) \right]\nonumber\\
=& \Phi \left( \frac{ \tsigma^2_V g\left( Z,\delta\right) \sigma_U \rho_{U}^\prime C^{-1}_\eta D^{-1}_\eta \eta}{\tilde{\sigma}_U(Z)\sigma^2(Z)}\right) f_{\varepsilon \vert \eta, 1} (\varepsilon \vert \eta ) + \Phi \left( - \frac{ \tsigma^2_V g\left( Z,\delta\right) \sigma_U \rho_{U}^\prime C^{-1}_\eta D^{-1}_\eta \eta}{\tilde{\sigma}_U(Z)\sigma^2(Z)}\right) f_{\varepsilon \vert \eta, 2} (\varepsilon \vert \eta ) ,
\end{align}
with
\[
\lambda(Z) = \frac{\tilde{\sigma}_U(Z)}{\tilde{\sigma}_V}, \text{ and }\sigma^2(Z) = \tilde{\sigma}^2_V + \tilde{\sigma}^2_U(Z),
\]
and $\Phi$ the cdf of a standard normal distribution. The distribution of $\varepsilon$ given $\eta$ is a mixture of two conditional extended skew-normal distributions, where the mixing probabilities depend on $\rho_U$ \citep[see][p. 35-36]{azzalini2013}. When $\rho_U=\Z$, that is, all the elements of $\rho_U$ are equal to zero, the mixing probabilities are both equal to $0.5$, and the conditional distribution of $\varepsilon$ reduces to a skew-normal distribution. That is, our specification reduces to a stochastic frontier model where the regressors are independent of the inefficiency term $U_0$.

The full information likelihood function is therefore given by
\[
\mathcal{L}(\theta) = f_{\varepsilon \vert \eta} (\varepsilon \vert \eta ; \beta, \delta, \sigma^2_V,\sigma^2_U,\rho_V,\rho_U) f_{\eta}(\eta; \gamma, diag(D_\eta), ve(C_\eta) ),
\]
where $\theta = (\beta^\prime,\delta^\prime,\sigma^2_V,\sigma^2_U,\rho^\prime_V,\rho^\prime_U,\gamma^\prime,diag(D_\eta)^\prime, ve(C_\eta)^\prime)^\prime$; $diag(D_\eta)$ denotes the diagonal of the matrix $D_\eta$; and $ve(\cdot)$ denotes the half-vectorization of the matrix $C_\eta$ which only keeps the $(p+k)(p+k-1)/2$ elements below the main diagonal (as the matrix is symmetric and the elements of the main diagonal are equal to $1$ by construction).\footnote{This operation is defined more formally as $ve(C_\eta) = L vech(C_\eta)$, where $L$ is an elimination matrix of dimension $(p+k)(p+k-1)/2 \times (p+k)(p+k+1)/2$, which only keeps the off-diagonal elements of the half-vectorization of the matrix $C_\eta$.} 

\subsection{Identification}

Let $\ell(\theta) = \log \mathcal{L}(\theta)$ be the log-likelihood function, and assume that $E \left[ \vert \ell(\theta) \vert \right] < \infty$ for all $\theta \in \Theta$. We define
\begin{equation} \label{eq:truepar2}
\theta_0 = \argmax_{\theta \in \Theta} E \left[ \ell(\theta) \right].
\end{equation}
As we can restrict $\Theta$ to be a compact parameter space, and the likelihood function is continuous in $\theta$, there exists a parameter vector $\theta_0$ which satisfies \eqref{eq:truepar2} \citep{gourieroux1990}.

We focus our identification analysis on the parameter $\rho_U$. To this end, we maintain the following assumption.
\begin{assumption} \label{ass:sigmau0pos}
Let $\theta_1 = (\beta^\prime,\delta^\prime,\sigma^2_V,\sigma^2_U,\rho^\prime_V,\gamma^\prime,diag(D_\eta)^\prime, ve(C_\eta)^\prime)^\prime$. The matrix
\[
E \left[ \nabla_{\theta_{1}\theta^\prime_{1}}^2 \ell(\theta_0) \right]
\]
is negative definite and has full rank. 
\end{assumption}

This assumption imposes that the parameter $\theta_1$ is first-order locally identified \citep{sargan1983}. In particular, we require that the variance of the inefficiency term $\sigma^2_{U,0} > 0$. \citet{lee1986} and \citet{lee1993} have shown that when $\sigma^2_{U,0} = 0$, the stochastic frontier model is not first-order identified. Moreover, in our model, whenever $\sigma^2_U = 0$, $(\delta,\rho_U)$ are not identified. We believe this case is worthy of future investigation, but we rule it out here for simplicity.

\begin{proposition} \label{prop:identif1}
Let Assumptions \ref{ass:cindepeta}-\ref{ass:sigmau0pos} hold, and $\rho_{U,0}$ to be such that
\[
E \left[ \nabla_{\rho_U} \ell(\theta_{1,0},\rho_{U,0}) \right] = 0.
\]
We have that
\begin{itemize}
\item[(i)] $E \left[ \nabla_{\rho_U} \ell(\theta_{1,0},-\rho_{U,0}) \right] = 0$.
\item[(ii)] $\nabla_{\rho_U} \ell(\theta_{1},\Z) = 0$, for any $\theta_1$.
\item[(iii)] If $\rho_{U,0}$ has at least one non-zero component, then the model is first-order identified.
\end{itemize}
\end{proposition}

Part (i) states that if $\rho_{U,0}$ is a solution of the maximization problem in \eqref{eq:truepar2}, so is $-\rho_{U,0}$, where the negative sign is applied to all components of the vector $\rho_{U,0}$. That is, the sign of all components of $\rho_U$ is not identified. Part (ii) implies that $\rho_{U,0} = \Z$ is always a solution of \eqref{eq:truepar2}, which is true for any value of $\theta_1$. This result entails that the matrix of second derivatives has rank equal to $dim(\theta) - p - k$, and the model is not first-order identified at $\rho_U = \Z$. In Part (iii), we show that first-order identification is restored when at least one component of $\rho_{U,0}$ is non-zero. A proof of this Proposition is provided in Appendix \ref{sec:appA}.\footnote{A similar identification problem arises in Zero Inefficiency Stochastic Frontier models, see \citet{kumbhakar2013,rho2015}.}

\begin{figure}[!ht]
  \begin{center}
    \includegraphics[scale=.5,trim={1in 2in 1in 1.5in}]{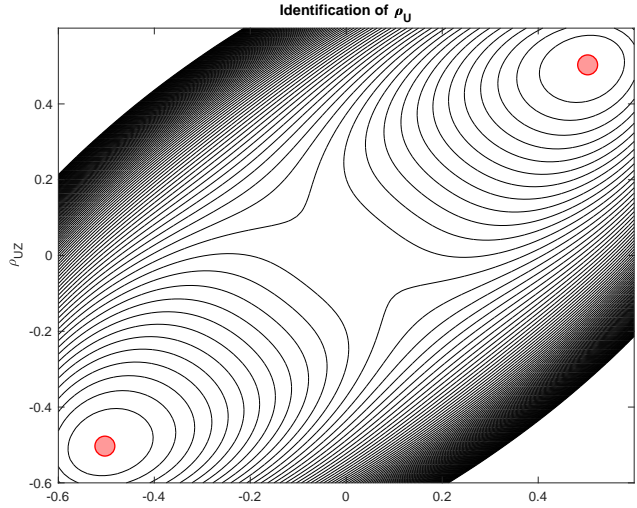}
    \caption{\label{fig:rhoident}Example of lack of identification of the parameter $\rho_U$.}
  \end{center}
\end{figure}

Figure \ref{fig:rhoident} illustrates the result of Part (i) of Proposition \ref{prop:identif1}. In this example, there are two endogenous regressors, one among the inputs and one among the environmental variables, so that $p = k = 1$, and the true value of $\rho_U = (0.5,0.5)^\prime$. The solid black lines are the level curves of the log-likelihood as a function of $\rho_U$, when all other parameters are taken to be known. The red dots designate the points where the log-likelihood function reaches its maximum. We can observe how both $(-0.5,-0.5)^\prime$ and $(0.5,0.5)^\prime$ are maxima of the log-likelihood function. Moreover, it can be seen from the level curves that, in this case, the log-likelihood also has a local minimum at $\rho_U = (0,0)^\prime$. 

We deal with the lack of identification of the sign of $\rho_U$ by restricting the support of one of its components, say the first one, $\rho_{U1}$, to be $[0,1]$ \citep{sundberg1974}. The sign of all other components is identified relative to this normalization, and provided $\rho_{U1,0}$ is in the interior of $[0,1]$.\footnote{An alternative approach would be to construct confidence sets for the identified set following \citet{chen2018}, but we do not pursue it here.} When $\rho_{U1} = 0$, the sign of the other components of $\rho_U$ remains unidentified. We can therefore distinguish two cases. In the first case, there is at least one component of $\rho_U$ which is non-zero. That is, $\rho_{U1,0} > 0$, and all other components are in the interior of $[-1,1]^{p + k - 1}$. In this case, the model is first order identified, and all the components of $\rho_U$ are identified up to a sign normalization (see Proposition \ref{prop:identif1}(iii)). In the second case, $\rho_{U} = \Z$, and, because of Proposition \ref{prop:identif1}(ii), the model is not first-order identified. We refer interested readers to Section \ref{sec:empiricalapp}, where we informally discuss the choice of $\rho_{U1}$ in practice.

We let $\bar{\Theta}$ to be the parameter's space which embeds the restriction on $\rho_{U1}$, and we redefine
\begin{equation} \label{eq:truepar}
\theta_0 = \argmax_{\theta \in \bar{\Theta}} E \left[ \ell(\theta) \right],
\end{equation}
which exists and is (locally) unique under Assumption \ref{ass:sigmau0pos}. 

While Part (ii) of Proposition \ref{prop:identif1} states that the model is not first-order identified when $\rho_{U,0} = \Z$, in the next proposition we show that the model is second-order identified at $\rho_{U,0} = \Z$.

\begin{proposition} \label{prop:identif2}
Let Assumption \ref{ass:cindepeta}-\ref{ass:sigmau0pos} hold, with $\rho_{U,0} = \Z$. Then
\[
\nabla_{\rho_U \rho^\prime_U}  \ell(\theta_{1,0},\Z)
\]
is not identically equal to $\Z$, and 
\[
E \left[ \nabla_{\rho_U \rho^\prime_U}  \ell(\theta_{1,0},\Z) \right] =\Z. 
\]
\end{proposition}

A proof is provided in Appendix \ref{sec:appA}.

\subsection{Estimation and Inference} \label{sec:est}

We consider an iid sample drawn from the joint distribution of $(Y,X,Z,W)$, that we denote $\lbrace (Y_i,X_i,Z_i,W_i),i = 1,\dots,n\rbrace$, where each observation follows the model in equation \eqref{stmod2}.

Estimation is straightforward and follows from the specification of the likelihood function derived above. For all $i = 1,\dots,n$, we can write
\begin{equation} \label{eq:fullloglik}
\mathcal{L}_n(\theta) = \prod_{i= 1}^n  f_{\varepsilon \vert \eta} (\varepsilon_i \vert \eta_i ; \beta, \delta, \sigma^2_V,\sigma^2_U,\rho_V,\rho_U) f_{\eta}(\eta_i; \gamma, diag(D_\eta), ve(C_\eta) ),
\end{equation}
with $\eta_i = (\eta^\prime_{X,i},\eta^\prime_{Z,i})^\prime$ and 
\begin{align*}
\varepsilon_i =& Y_i - m(X_i,\beta)\\
\eta_{Xi} =& X_i - W_i \gamma_X\\
\eta_{Zi} =& Z_i - W_i \gamma_Z.
\end{align*}

Letting, $\ell_n(\theta) = \log \mathcal{L}_n(\theta)$ to be the sample log-likelihood function, we have
\[
\hat\theta_n = \argmax_{\theta \in \bar{\Theta}} \ell_n(\theta).
\]
In parallel with our identification study, we analyze our estimator's asymptotic properties depending on the true value of the parameter $\rho_U$. 

Upon the additional assumption that $E\left[ \sup_{\theta \in \bar{\Theta}} \vert \ell(\theta) \vert \right] < \infty$, the log-likelihood function satisfies the required conditions for consistency \citep[see][Theorem 2.5, p. 2131]{newey1994h}. We thus have that
\[
\hat{\theta}_{n} \xrightarrow{p} \theta_{0}. 
\]
Moreover, the log-likelihood function is at least twice continuously differentiable with respect to the parameter $\theta_{0}$. When $\rho_U$ is in the interior of $[0,1] \times [-1,1]^{p+k-1}$ and Assumption \ref{ass:sigmau0pos} holds, standard theory of maximum likelihood estimation applies, and we can claim that 
\[
\sqrt{n} \left( \hat{\theta}_{n} - \theta_{0} \right) \xrightarrow{d} N\left(0,\mathcal{I}_{\theta_{0}}^{-1}  \right),
\]
where $\mathcal{I}_{\theta_{0}}$ is the Fisher's information matrix.

However, the asymptotic distribution and the rate of convergence of our estimator are non-standard when all the components of $\rho_{U,0}$ are equal to $0$. In this case, it follows from the result of Proposition \ref{prop:identif1} that we have a singular Hessian matrix, and one of the parameters of interest is at the boundary of the parameter space. This implies that we do not have the standard $\sqrt{n}$-rate of convergence, and that our estimator is not asymptotically normal \citep{sundberg1974b,andrews1999,rot2000}. However, Proposition \ref{prop:identif2} also implies that a reparametrization of the log-likelihood function allows us to obtain the rate of convergence and asymptotic distribution of our estimator.

Let $vec(\rho_U \rho_U^\prime)$ be the $(p+k)^2$ vectorization of the matrix $\rho_U \rho_U^\prime$. The following theorem gives the asymptotic properties of our estimator when $\rho_{U,0} = \Z$.

% HERE YOU JUST HAVE TO BORROW THE NOTATIONS AND THEOREMS OF ANDREWS (1999) AND SPECIALIZE THEM TO THIS CASE.
\begin{theorem}\label{thm:asnorm}
Let Assumptions \ref{ass:cindepeta}-\ref{ass:sigmau0pos} hold with $\rho_{U,0} = \Z$, and $(Z_{\theta_1},Z_{\rho_U\rho^\prime_U})$ a normal random vector such that $dim(Z_{\theta_1}) = dim(\theta_{1})$, $dim(Z_{\rho_U\rho^\prime_U}) = (p+k)^2$, with covariance matrix $\mathcal{I}_1^{-}$, where $\mathcal{I}_1^{-}$ is the pseudo-inverse of
\[
\mathcal{I}_1 = \begin{bmatrix} \mathcal{I}_{\theta_1} & \mathcal{I}_{\theta_1\rho_U\rho^\prime_U} \\ \mathcal{I}_{\rho_U\rho^\prime_U\theta_1} & \mathcal{I}_{\rho_U\rho^\prime_U} \end{bmatrix}.
\]
Define the quadratic function
\[
Q_1(\tau) = \left(\tau - Z_{\rho_U\rho^\prime_U} \right)^\prime \left[ \mathcal{I}_{\rho_U\rho^\prime_U}  - \mathcal{I}_{\rho_U\rho^\prime_U\theta_1} \mathcal{I}^{-1}_{\theta_1} \mathcal{I}_{\theta_1\rho_U\rho^\prime_U}\right]\left(\tau - Z_{\rho_U\rho^\prime_U} \right),
\]
and $\hat{\tau}_{\rho_U\rho^\prime_U}$ such that 
\[
Q_1(\hat\tau_{\rho_U\rho^\prime_U}) = \inf_{\tau \in \Tau_1} Q_1(\tau),
\]
where $\Tau_1 = \lbrace \tau \in \R^{(p+k)^2}: \tau \geq 0 \rbrace$. Then
\begin{itemize}
\item[(i)]
\[
\sqrt{n}\begin{pmatrix} \hat{\theta}_1 - \theta_{1,0}\\ vech(\hat\rho_U \hat\rho_U^\prime) \end{pmatrix} \xrightarrow{d} \begin{pmatrix} Z_{\theta_1} - \mathcal{I}_{\theta_1}^{-1} \mathcal{I}_{\theta_1\rho_U \rho_U^\prime} \hat\tau_{\rho_U\rho^\prime_U} \\ \hat\tau_{\rho_U\rho^\prime_U} \end{pmatrix} 
\]
\item[(ii)] $n^{1/4} \hat\rho_U = O_P(1)$. 
\end{itemize}
\end{theorem}

The vector $\hat\tau_{\rho_U\rho^\prime_U}$ is the projection of a normal random vector onto $\Tau_1$ with respect to the Euclidean norm weighted by the matrix $\mathcal{I}_{\rho_U\rho^\prime_U}  - \mathcal{I}_{\rho_U\rho^\prime_U\theta_1} \mathcal{I}^{-1}_{\theta_1} \mathcal{I}_{\theta_1\rho_U\rho^\prime_U}$ \citep{chernoff1954,andrews1999,rot2000}. When $\rho_U$ is a scalar, $\hat\tau_{\rho_U\rho^\prime_U} = \max \lbrace Z_{\rho_U \rho^\prime_U},0\rbrace$. However, it is more cumbersome to derive the distribution of $\hat\tau_{\rho_U\rho^\prime_U}$ in closed form when the dimension of $\rho_U$ is greater than one and when there is dependence between the components of $vec(\rho_U\rho^\prime_U)$.\footnote{We show in a Supplementary Appendix that the off-diagonal elements of the $(p+k)^2 \times (p+k)^2$ matrix of fourth derivatives wrt $\rho_U$ are not zero in general.} The result in Part (ii) is a direct consequence of Part (i). However, it is worth highlighting that our estimator has rates of convergence slower than $\sqrt{n}$, and may not be asymptotically normal, depending on the true value of the parameter $\rho_U$. These results have important implications for obtaining standard errors and conducting inference on $\rho_U$. 

To obtain standard errors and confidence intervals, we advocate the use of the subsampling method of \citet{politis1994}; or the $m$-out-of-$n$ bootstrap of \citet{andrews2000}. The subsampling method of \citet{politis1994} is consistent whenever the estimator has some asymptotic distributions (not necessarily normal) and when the rate of convergence is slower than $\sqrt{n}$.\footnote{One potential issue with the subsampling method is that one has to know the rate of convergence of the estimator. In practice, one can test first whether $\rho_U = \Z$, and then apply the appropriate rate of convergence. Also, \citet{bertail1999} extend the subsampling method to the case when rates of convergence are unknown. We do not explore it here.} \citet{andrews2000} has shown that the $m$-out-of-$n$ bootstrap is consistent when the true parameter is at the boundaries but only when the estimator is $\sqrt{n}$ convergent. Provided that $m^2/n = o(1)$, the $m$-out-of-$n$ bootstrap is consistent when rates of convergence are slower than $\sqrt{n}$ \citep{bertail1999}. We refer interested readers to \citet{andrews2010} for a recent study of asymptotic uniformity of subsampling and of the m-out-of-n bootstrap.

Furthermore, one may wish to conduct inference on the parameter $\rho_U$. In particular, a simple hypothesis to be tested is whether $X$ and $Z$ are independent of the inefficiency term, i.e. $\rho_U =\Z$. The trinity of tests is an obvious candidate, but the implementation of these tests is not straightforward because of the non-standard asymptotic properties of the estimator of $\rho_U$. 

\citet{andrews2001} studies the properties of the trinity of test when some parameters are at the boundary, although the author does not consider the issue of singularity of the Hessian matrix. His theoretical results about the Likelihood Ratio (LR) test can nonetheless be used following Theorem \ref{thm:asnorm}, provided one can obtain an estimator of the information matrix under $H_0: vec(\rho_U \rho^\prime_U)= \Z$. The critical values from the asymptotic distribution of the LR statistic are obtained by random draws from the vector $(Z_{\theta_1},Z_{\rho_U \rho^\prime_U})$, and by solving a quadratic programming problem \citep[see][]{andrews1999,andrews2001}.

%Similarly, Wald, Score, and LR test and their modifications developed in \citet{andrews2001} could be used under the result of our Theorem \ref{thm:asnorm2}.

One important remark is about the Score test. Irrespective of the true value of $\rho_U$, the Score test has no power around $\rho_U=\Z$. This is because, as shown in Proposition \ref{prop:identif1}, the score is always identically zero at that point. 

We leave a thorough theoretical exploration of the properties of the Trinity of tests in this model for future work, but we explore some of the finite sample properties of the LR test in simulations.

\subsection{Technical Efficiency} Our framework is completed by an estimator of technical efficiency, $TE = \exp(-U_i)$, which is obtained from the conditional distribution of $U$ given $\varepsilon$ and $\eta$.

Let 
\begin{align*}
\lambda_{\star} =& \sqrt{1 + \lambda^2(Z)}\\
\sigma_{\star} =& \frac{\tilde\sigma_V \tilde\sigma_U(Z)}{\sigma(Z)} \\
\mu_{1\star} =& -\left( \varepsilon - \sigma_V \rho_{V}^\prime C^{-1}_\eta D^{-1}_\eta \eta \right) \frac{\tilde\sigma^2_U(Z)}{\sigma^2(Z)} \\
\mu_{2\star} =& g\left( Z,\delta\right)\sigma_U \rho_{U}^\prime C^{-1}_\eta D^{-1}_\eta \eta \frac{\tilde\sigma^2_V}{\sigma^2(Z)},
\end{align*}
where we have removed the dependence of $\lambda_{\star}$, $\sigma_{\star}$, $\mu_{1\star}$ and $\mu_{2\star}$ on $Z$ for simplicity. By equation \eqref{eq:conddens}, we have that the joint density of $(\varepsilon,\eta)$ can be written as 
\begin{align*}
f_{\varepsilon,\eta} (\varepsilon,\eta) =& \left( \Phi \left( \frac{\mu_{2\star}}{\sigma_{\star} \lambda_{\star}} \right) f_{\varepsilon \vert \eta, 1} (\varepsilon \vert \eta ) + \Phi \left( -\frac{\mu_{2\star}}{\sigma_{\star} \lambda_{\star}}\right) f_{\varepsilon \vert \eta, 2} (\varepsilon \vert \eta )\right) f_{\eta} (\eta)\\
=&\Phi \left( \frac{\mu_{2\star} }{\sigma_{\star} \lambda_{\star}}\right) f_{\varepsilon, \eta, 1} (\varepsilon, \eta ) + \Phi \left( -\frac{\mu_{2\star}}{\sigma_{\star}\lambda_{\star}} \right)f_{\varepsilon, \eta, 2} (\varepsilon, \eta ).
\end{align*}
The conditional density of $U$ given $\varepsilon$ and $\eta$ is then equal to
\begin{align*}
f_{U\vert\varepsilon,\eta}  (u \vert \varepsilon ,\eta )  =& \frac{1}{\sqrt{2\pi} \sigma_{\star}}\left\lbrace  \Phi \left( \frac{\mu_{2\star}}{\sigma_{\star}\lambda_{\star}}\right) \frac{f_{\varepsilon, \eta, 1} (\varepsilon, \eta ) }{f_{\varepsilon,\eta} (\varepsilon,\eta)} \left[ \Phi\left( \frac{\mu_{1\star} + \mu_{2 \star}}{\sigma_{\star}}\right)\right]^{-1} \exp \left( -\frac{\left(u - \mu_{1\star}  - \mu_{2 \star}\right)^2}{2\sigma^2_{\star}}\right) \right. \\
&\quad \left. + \Phi \left( -\frac{\mu_{2\star}}{\sigma_{\star}\lambda_{\star}}\right)\frac{f_{\varepsilon, \eta, 2} (\varepsilon, \eta ) }{f_{\varepsilon,\eta} (\varepsilon,\eta)}\left[ \Phi\left( \frac{\mu_{1\star} - \mu_{2 \star}}{\sigma_{\star}}\right)\right]^{-1}  \exp \left( -\frac{\left(u - \mu_{1\star} + \mu_{2\star} \right)^2}{2\sigma^2_{\star}}\right) \right\rbrace.
\end{align*}
When both $U_0$ and $V$ are independent of $\eta$, this conditional density reduces to the one derived in \citet{jondrow1982}. 

Hence
\begin{align*}
E \left[ \exp(-U) \vert \varepsilon, \eta \right] =& \frac{1}{\sqrt{2\pi}\sigma_{\star}}\left\lbrace  \Phi \left( \frac{\mu_{2\star}}{\sigma_{\star}\lambda_{\star}} \right)\frac{f_{\varepsilon, \eta, 1} (\varepsilon, \eta ) }{f_{\varepsilon,\eta} (\varepsilon,\eta)} \left[ \Phi\left( \frac{\mu_{1\star} + \mu_{2 \star}}{\sigma_{\star}}\right)\right]^{-1} \int_{0}^\infty \exp \left( -u - \frac{\left( u -  \mu_{1\star} - \mu_{2\star}\right)^2}{2\sigma^2_{\star}}\right) du \right. \\
&\quad \left. +  \Phi \left( -\frac{\mu_{2\star}}{\sigma_{\star}\lambda_{\star}}\right) \frac{f_{\varepsilon, \eta, 2} (\varepsilon, \eta ) }{f_{\varepsilon,\eta} (\varepsilon,\eta)} \left[ \Phi\left( \frac{\mu_{1\star} - \mu_{2 \star}}{\sigma_{\star}}\right)\right]^{-1} \int_{0}^\infty \exp \left( -u - \frac{\left( u -  \mu_{1\star} + \mu_{2\star}\right)^2}{2\sigma^2_{\star}}\right) du \right\rbrace.
\end{align*}
By the properties of the cdf of the univariate normal distribution, this expression is shown to be equal to 
\begin{align}
E \left[ \exp(-U) \vert \varepsilon, \eta \right] =&  \Phi \left( \frac{\mu_{2\star}}{\sigma_{\star}\lambda_{\star}}\right) \frac{f_{\varepsilon, \eta, 1} (\varepsilon, \eta ) }{f_{\varepsilon,\eta} (\varepsilon,\eta)} \exp \left(-\mu_{1\star} - \mu_{2\star} + \frac{\sigma^2_{\star}}{2}\right)\frac{1 - \Phi \left( - \frac{\mu_{1\star} + \mu_{2\star}}{\sigma_{\star}}  + \sigma_{\star}\right)}{\Phi\left( \frac{\mu_{1\star} + \mu_{2 \star}}{\sigma_{\star}}\right)}  \nonumber \\
& \quad +  \Phi \left( -\frac{\mu_{2\star}}{\sigma_{\star}\lambda_{\star}}\right) \frac{f_{\varepsilon, \eta, 2} (\varepsilon, \eta ) }{f_{\varepsilon,\eta} (\varepsilon,\eta)}\exp \left(-\mu_{1\star} + \mu_{2\star} + \frac{\sigma^2_{\star}}{2}\right) \frac{1 - \Phi \left( -\frac{\mu_{1\star} - \mu_{2\star}}{\sigma_{\star}}  + \sigma_{\star}\right)}{\Phi\left( \frac{\mu_{1\star} - \mu_{2 \star}}{\sigma_{\star}}\right)} \label{eq:battesecoeli}.
\end{align}

This formula generalizes \citet{battese1988} formula for technical efficiencies to the endogenous case. Finally, the mean technical efficiency can be obtained as
\[
E \left[ \exp(-U)  \right] = E \left[E \left[ \exp(-U) \vert \varepsilon, \eta \right] \right],
\]
by the law of iterated expectations \citep{lee1978}.

\section{Simulations} \label{sec:montecarlo}

We replicate the simulation scheme in \citet{amsler2017}. We consider the following model
\[
Y_i = \beta_0 + X_{1i} \beta_1 + X_{2i} \beta_2 + V_i - U_{0i} \exp\left( Z_{1i} \delta_1 + Z_{2i} \delta_2\right),
\]
with $\beta_0 =\delta_1 = \delta_2 =0$ and $\beta_1 = \beta_2 =0.661$, and where the random variables $(X_{1i},Z_{1i})$ are taken to be exogenous (i.e. independent of the composite error term), and $(X_{2i},Z_{2i})$ are instead endogenous. We consider two instruments $(W_{1i},W_{2i})$, also independent of the error term.

The exogenous variables are generated independently from a normal distribution with means equal to $0$ and variances equal to $1$. These variables are equicorrelated, with correlation parameter equal to $0.5$. 

We generate the triplet $(V,\eta_X,\eta_Z)$ from the following normal distribution
\[
\begin{pmatrix} V_i \\ \eta_{X,i} \\ \eta_{Z,i} \end{pmatrix} \sim N \left( \begin{bmatrix} 0 \\ 0 \\ 0 \end{bmatrix},\begin{bmatrix} 1 & \rho^\prime_V  \\ \rho_V & \Sigma_\eta  \end{bmatrix}\right), 
\]
with $\rho_V = (0.5,0.5)^\prime$,
\[
\Sigma_\eta = C_\eta= \begin{bmatrix} 1  & \rho_{\eta,12}   \\ \rho_{\eta,12}  & 1 \end{bmatrix} = \begin{bmatrix} 1 & 0.5  \\ 0.5 & 1 \end{bmatrix},
\]
and
\begin{align*}
X_{2i} =& \gamma \left( X_{1i} + Z_{1i} + W_{1i} + W_{2i} \right) + \eta_{X,i} \\
Z_{2i} =& \gamma \left( X_{1i} + Z_{1i} + W_{1i} + W_{2i} \right) + \eta_{Z,i}, 
\end{align*}
with $\gamma = 0.316$. 

We finally generate
\[
U^\ast_0 \sim N \left( \sigma_U \rho^\prime_U C^{-1}_{\eta} \eta, \sigma^2_U (1 - \rho^\prime_U C^{-1}_{\eta} \rho_U) \right),
\]
with the stochastic inefficiency term given by $U_0 = \vert U^\ast_0 \vert$. 

We consider two simulation schemes that differ because of the value of the parameter $\rho_U$. In \textit{Setting 1}, we take $U_0$ to be independent of $\eta$ \citep[the same setting as in][]{amsler2017}. In \textit{Setting 2}, we take $\rho_U = (0.5,0.5)^\prime$. In both settings, we impose that the first component of $\rho_U$ belongs to $[0,1]$. We take increasing sample sizes $n =\lbrace 250,500,1000 \rbrace$, and run $1000$ replications for each scenario. 

Our estimation procedure is based on the maximization of the full likelihood in equation \eqref{eq:fullloglik}. 

There are two main issues for practical implementation of these models. First, the parameter space is often very large. To reduce the dimensionality of the optimization problem, one can first estimate the vector of parameters $\left( \gamma,diag(D_\eta),ve(C_\eta)\right)$ by OLS. Given $\left( \gamma,diag(D_\eta),ve(C_\eta)\right)$, one can then maximize the full likelihood with respect to the other parameters.

Moreover, the starting values for the remaining parameters need to be appropriately chosen, especially in nonlinear, high dimensional optimization problems like ours. To this end, we use the method of moments. We can write 
\[
E\left[ Y_i \vert X_i,Z_i,\eta_i\right] = \beta_0 + X_{1i} \beta_1 + X_{2i} \beta_2 + E\left[ V_i \vert \eta_i \right] - E\left[ U_{0i} \vert \eta_i \right]\exp\left( Z_{1i} \delta_1 + Z_{2i} \delta_2\right),
\]
using the assumption that $(U_0,V)$ is independent of $(X_2,Z_2)$ given $\eta$, with
\begin{align*}
E\left[ V_i \vert \eta_i \right] =& \sigma_V \rho^\prime_{V} C_\eta^{-1} D^{-1}_\eta\eta_i\\
E\left[ U_{0i} \vert \eta_i \right] =& 2\sigma_U\sqrt{1 -  \rho_{U}^\prime C^{-1}_\eta \rho_{U} }\phi\left( \frac{\rho^\prime_{U} C_\eta^{-1} D^{-1}_\eta\eta_i}{\sqrt{ 1 -  \rho_{U}^\prime C^{-1}_\eta \rho_{U}}}\right) \\
&\quad + \left(2\Phi\left( \frac{\rho^\prime_{U} C_\eta^{-1} D^{-1}_\eta\eta_i}{\sqrt{ 1 -  \rho_{U}^\prime C^{-1}_\eta \rho_{U}}}\right) - 1\right)\sigma_U \rho^\prime_{U} C_\eta^{-1} D^{-1}_\eta\eta_i.
\end{align*}

We report both the average standard errors obtained by evaluating numerically the Hessian matrix of the full likelihood (Av. SE), the coverage of Wald-type confidence intervals (CI), and the coverage of confidence intervals obtained by the random subsampling method of \citet{politis1994} (CI$^\ast$). The nominal size for both is equal to $95\%$. The size of each subsample, $b$, should be selected in such a way that $b \rightarrow \infty$ and $b/n = o(1)$. We use $b = \lfloor n^{0.95}/\log(n)\rfloor$, where $\lfloor \cdot \rfloor$ denotes the integer part of a number. Choosing the size of each subsample in a data-driven way is still an open question and we do not explore it here \citep[see][]{politis1999}. Based on our theoretical results, we expect that the inversion of the Hessian matrix is not going to provide reliable estimates of the standard errors in \textit{Setting 1}.

Tables \ref{table:mc1} and \ref{table:mc2} below contain the results of these simulations. Table \ref{table:mc1} should be compared with Table 4, p. 138 of \citet{amsler2017}. The mean and the standard deviation for most of the parameters are comparable with theirs. However, we achieve much better precision in estimating the variance of the inefficiency term, which, as indicated by \citet{amsler2017}, is estimated imprecisely using the copula method. Both the bias and the variance decrease as the sample size $n$ increases, which ought to be expected from our MLE. The average standard errors computed using the inverse of the numerical Hessian are generally larger than the sampling standard deviation. However, Wald-type confidence intervals have good coverage, with the exception of those for the dependent parameter $\rho_U$ whose coverage is well below the nominal one. Subsampling confidence intervals have good coverage for the parameters of the stochastic frontier model. However, they undercover the first stage parameters, especially the variances of the control functions.

\begin{table}[hb]
\centering
\adjustbox{max width=\textwidth}{%
\scriptsize
\begin{tabular}{l l | c c c c c |  c c c c c | c c c c c} \hline \hline
~ & ~ & \multicolumn{4}{c}{$N = 250$} & \multicolumn{4}{c}{$N = 500$} & \multicolumn{4}{c}{$N = 1000$}\\ \hline
~ & TRUE & Mean & Std. Dev. & Av. SE & CI & CI$^\ast$ & Mean & Std. Dev. & Av. SE & CI & CI$^\ast$ & Mean & Std. Dev. & Av. SE & CI & CI$^\ast$\\ \hline
% latex table generated in R 4.0.4 by xtable 1.8-4 package
% Thu Mar 18 00:30:41 2021
  \hline
$\beta_0$ & 0.000 & -0.151 & 0.420 & 0.253 & 0.828 & 0.857 & -0.067 & 0.288 & 0.162 & 0.886 & 0.925 & -0.022 & 0.149 & 0.093 & 0.923 & 0.953 \\ 
  $\beta_1$ & 0.661 & 0.666 & 0.148 & 0.264 & 0.931 & 0.974 & 0.659 & 0.102 & 0.158 & 0.928 & 0.977 & 0.665 & 0.074 & 0.074 & 0.909 & 0.949 \\ 
  $\beta_2$ & 0.661 & 0.654 & 0.174 & 0.292 & 0.911 & 0.976 & 0.661 & 0.115 & 0.142 & 0.930 & 0.982 & 0.657 & 0.080 & 0.084 & 0.924 & 0.974 \\ 
  $\delta_1$ & 0.000 & 0.027 & 0.426 & 0.606 & 0.904 & 0.945 & 0.018 & 0.265 & 1.462 & 0.936 & 0.977 & 0.004 & 0.131 & 0.051 & 0.948 & 0.994 \\ 
  $\delta_2$ & 0.000 & -0.004 & 0.275 & 0.420 & 0.943 & 0.957 & -0.019 & 0.191 & 1.074 & 0.945 & 0.973 & -0.007 & 0.080 & 0.039 & 0.956 & 0.992 \\ 
  $\sigma^2_{U}$ & 2.752 & 2.404 & 1.115 & 0.629 & 0.870 & 0.828 & 2.590 & 0.752 & 0.532 & 0.920 & 0.924 & 2.692 & 0.420 & 0.361 & 0.941 & 0.954 \\ 
  $\sigma^2_{V}$ & 1.000 & 1.114 & 0.418 & 0.406 & 0.801 & 0.827 & 1.052 & 0.267 & 0.239 & 0.872 & 0.908 & 1.020 & 0.164 & 0.139 & 0.884 & 0.935 \\ 
  $\rho_{U,\eta_X}$ & 0.000 & 0.149 & 0.222 & 0.496 & 0.786 & 0.854 & 0.113 & 0.160 & 0.293 & 0.793 & 0.949 & 0.100 & 0.100 & 0.155 & 0.754 & 0.993 \\ 
  $\rho_{U,\eta_Z}$ & 0.000 & 0.059 & 0.259 & 0.563 & 0.812 & 0.972 & 0.066 & 0.189 & 0.420 & 0.810 & 0.986 & 0.047 & 0.144 & 0.150 & 0.785 & 0.992 \\ 
  $\rho_{V,\eta_X}$ & 0.500 & 0.484 & 0.157 & 0.217 & 0.869 & 0.930 & 0.489 & 0.109 & 0.209 & 0.870 & 0.950 & 0.499 & 0.072 & 0.074 & 0.906 & 0.958 \\ 
  $\rho_{V,\eta_Z}$ & 0.500 & 0.489 & 0.112 & 0.118 & 0.856 & 0.931 & 0.493 & 0.081 & 0.130 & 0.891 & 0.949 & 0.497 & 0.050 & 0.051 & 0.924 & 0.965 \\ 
  $\gamma_{x,0}$ & 0.000 & -0.001 & 0.065 & 0.083 & 0.933 & 0.923 & -0.001 & 0.044 & 0.077 & 0.944 & 0.927 & 0.001 & 0.031 & 0.032 & 0.949 & 0.933 \\ 
  $\gamma_{x,1}$ & 0.316 & 0.314 & 0.081 & 0.140 & 0.940 & 0.943 & 0.317 & 0.056 & 0.113 & 0.942 & 0.938 & 0.318 & 0.040 & 0.041 & 0.938 & 0.939 \\ 
  $\gamma_{x,2}$ & 0.316 & 0.323 & 0.079 & 0.125 & 0.939 & 0.960 & 0.316 & 0.055 & 0.069 & 0.944 & 0.946 & 0.317 & 0.038 & 0.040 & 0.951 & 0.939 \\ 
  $\gamma_{x,3}$ & 0.316 & 0.313 & 0.078 & 0.105 & 0.942 & 0.949 & 0.318 & 0.053 & 0.075 & 0.949 & 0.961 & 0.318 & 0.039 & 0.040 & 0.943 & 0.944 \\ 
  $\gamma_{x,4}$ & 0.316 & 0.313 & 0.078 & 0.115 & 0.939 & 0.950 & 0.316 & 0.052 & 0.108 & 0.957 & 0.962 & 0.314 & 0.037 & 0.040 & 0.955 & 0.945 \\ 
  $\gamma_{z,0}$ & 0.000 & -0.001 & 0.065 & 0.085 & 0.941 & 0.926 & -0.001 & 0.045 & 0.049 & 0.949 & 0.922 & 0.002 & 0.032 & 0.033 & 0.941 & 0.922 \\ 
  $\gamma_{z,1}$ & 0.316 & 0.315 & 0.080 & 0.102 & 0.951 & 0.936 & 0.316 & 0.056 & 0.068 & 0.954 & 0.938 & 0.316 & 0.039 & 0.041 & 0.947 & 0.936 \\ 
  $\gamma_{z,2}$ & 0.316 & 0.319 & 0.077 & 0.102 & 0.950 & 0.952 & 0.316 & 0.054 & 0.101 & 0.948 & 0.932 & 0.318 & 0.037 & 0.040 & 0.961 & 0.938 \\ 
  $\gamma_{z,3}$ & 0.316 & 0.314 & 0.076 & 0.103 & 0.948 & 0.951 & 0.317 & 0.052 & 0.104 & 0.955 & 0.954 & 0.318 & 0.037 & 0.040 & 0.961 & 0.941 \\ 
  $\gamma_{z,4}$ & 0.316 & 0.317 & 0.075 & 0.101 & 0.952 & 0.951 & 0.317 & 0.053 & 0.089 & 0.954 & 0.937 & 0.314 & 0.036 & 0.040 & 0.952 & 0.941 \\ 
  $\sigma^2_{\eta_X}$ & 1.000 & 0.977 & 0.091 & 0.125 & 0.903 & 0.867 & 0.992 & 0.063 & 0.179 & 0.930 & 0.909 & 0.995 & 0.043 & 0.044 & 0.942 & 0.934 \\ 
  $\sigma^2_{\eta_Z}$ & 1.000 & 0.983 & 0.091 & 0.129 & 0.916 & 0.895 & 0.993 & 0.062 & 0.163 & 0.935 & 0.913 & 0.997 & 0.045 & 0.045 & 0.936 & 0.929 \\ 
  $\rho_{\eta_X \eta_Z}$ & 0.500 & 0.498 & 0.050 & 0.079 & 0.923 & 0.937 & 0.502 & 0.034 & 0.093 & 0.930 & 0.940 & 0.500 & 0.024 & 0.023 & 0.942 & 0.937 \\ 
   \hline

\hline 
\end{tabular}}
\caption{\textit{Simulation results for \textit{Setting 1}}}
\label{table:mc1}
\end{table}

Results in \textit{Setting 2} are comparable to the results obtained above. It is worth noticing that, in line with our theory, standard errors are now estimated more precisely using the inverse of the Hessian matrix, and the coverage of Wald-type confidence intervals is much closer to the nominal one, for all parameters of the model. Subsampling confidence intervals perform similarly as above.

\begin{table}[ht]
\centering
\adjustbox{max width=\textwidth}{%
\scriptsize
\begin{tabular}{l l | c c c c c |  c c c c c | c c c c c} \hline \hline
~ & ~ & \multicolumn{4}{c}{$N = 250$} & \multicolumn{4}{c}{$N = 500$} & \multicolumn{4}{c}{$N = 1000$}\\ \hline
~ & TRUE & Mean & Std. Dev. & Av. SE & CI & CI$^\ast$ & Mean & Std. Dev. & Av. SE & CI & CI$^\ast$ & Mean & Std. Dev. & Av. SE & CI & CI$^\ast$\\ \hline
% latex table generated in R 4.0.4 by xtable 1.8-4 package
% Thu Mar 18 00:31:51 2021
  \hline
$\beta_0$ & 0.000 & -0.146 & 0.362 & 0.204 & 0.815 & 0.833 & -0.039 & 0.196 & 0.162 & 0.885 & 0.937 & -0.011 & 0.105 & 0.089 & 0.916 & 0.952 \\ 
  $\beta_1$ & 0.661 & 0.666 & 0.143 & 0.184 & 0.929 & 0.970 & 0.661 & 0.105 & 0.139 & 0.909 & 0.961 & 0.666 & 0.069 & 0.071 & 0.931 & 0.959 \\ 
  $\beta_2$ & 0.661 & 0.652 & 0.175 & 0.198 & 0.895 & 0.969 & 0.660 & 0.115 & 0.137 & 0.915 & 0.978 & 0.657 & 0.077 & 0.080 & 0.912 & 0.965 \\ 
  $\delta_1$ & 0.000 & 0.012 & 0.237 & 0.275 & 0.905 & 0.973 & -0.001 & 0.071 & 0.124 & 0.950 & 0.999 & -0.000 & 0.040 & 0.041 & 0.941 & 0.995 \\ 
  $\delta_2$ & 0.000 & -0.012 & 0.206 & 0.118 & 0.895 & 0.972 & -0.001 & 0.066 & 0.067 & 0.935 & 0.997 & -0.001 & 0.030 & 0.030 & 0.946 & 0.998 \\ 
  $\sigma^2_{U}$ & 2.752 & 2.353 & 1.072 & 0.628 & 0.831 & 0.834 & 2.634 & 0.633 & 0.516 & 0.902 & 0.940 & 2.713 & 0.365 & 0.348 & 0.932 & 0.961 \\ 
  $\sigma^2_{V}$ & 1.000 & 1.139 & 0.395 & 0.331 & 0.788 & 0.818 & 1.035 & 0.244 & 0.232 & 0.855 & 0.921 & 1.007 & 0.151 & 0.134 & 0.881 & 0.939 \\ 
  $\rho_{U,\eta_X}$ & 0.500 & 0.543 & 0.156 & 0.304 & 0.835 & 0.896 & 0.509 & 0.085 & 0.092 & 0.915 & 0.959 & 0.504 & 0.050 & 0.045 & 0.935 & 0.971 \\ 
  $\rho_{U,\eta_Z}$ & 0.500 & 0.518 & 0.170 & 0.212 & 0.854 & 0.954 & 0.507 & 0.082 & 0.080 & 0.926 & 0.987 & 0.501 & 0.049 & 0.045 & 0.929 & 0.994 \\ 
  $\rho_{V,\eta_X}$ & 0.500 & 0.478 & 0.163 & 0.179 & 0.845 & 0.907 & 0.490 & 0.107 & 0.098 & 0.880 & 0.945 & 0.500 & 0.071 & 0.072 & 0.912 & 0.961 \\ 
  $\rho_{V,\eta_Z}$ & 0.500 & 0.480 & 0.129 & 0.135 & 0.857 & 0.910 & 0.497 & 0.082 & 0.073 & 0.901 & 0.958 & 0.497 & 0.052 & 0.052 & 0.924 & 0.965 \\ 
  $\gamma_{x,0}$ & 0.000 & -0.000 & 0.065 & 0.070 & 0.931 & 0.936 & -0.002 & 0.043 & 0.051 & 0.944 & 0.928 & 0.001 & 0.031 & 0.032 & 0.947 & 0.927 \\ 
  $\gamma_{x,1}$ & 0.316 & 0.314 & 0.077 & 0.089 & 0.943 & 0.957 & 0.318 & 0.055 & 0.061 & 0.948 & 0.941 & 0.319 & 0.038 & 0.039 & 0.934 & 0.932 \\ 
  $\gamma_{x,2}$ & 0.316 & 0.323 & 0.077 & 0.082 & 0.944 & 0.958 & 0.315 & 0.052 & 0.059 & 0.940 & 0.943 & 0.316 & 0.037 & 0.038 & 0.951 & 0.942 \\ 
  $\gamma_{x,3}$ & 0.316 & 0.314 & 0.076 & 0.083 & 0.940 & 0.962 & 0.320 & 0.052 & 0.066 & 0.948 & 0.952 & 0.317 & 0.038 & 0.038 & 0.944 & 0.939 \\ 
  $\gamma_{x,4}$ & 0.316 & 0.314 & 0.074 & 0.089 & 0.947 & 0.953 & 0.315 & 0.051 & 0.058 & 0.950 & 0.957 & 0.315 & 0.037 & 0.037 & 0.937 & 0.946 \\ 
  $\gamma_{z,0}$ & 0.000 & 0.000 & 0.064 & 0.068 & 0.939 & 0.940 & -0.001 & 0.045 & 0.055 & 0.950 & 0.933 & 0.002 & 0.032 & 0.032 & 0.944 & 0.921 \\ 
  $\gamma_{z,1}$ & 0.316 & 0.315 & 0.077 & 0.091 & 0.952 & 0.951 & 0.317 & 0.055 & 0.060 & 0.952 & 0.936 & 0.317 & 0.037 & 0.039 & 0.956 & 0.932 \\ 
  $\gamma_{z,2}$ & 0.316 & 0.319 & 0.076 & 0.098 & 0.944 & 0.962 & 0.315 & 0.052 & 0.057 & 0.953 & 0.936 & 0.317 & 0.036 & 0.038 & 0.955 & 0.937 \\ 
  $\gamma_{z,3}$ & 0.316 & 0.315 & 0.075 & 0.091 & 0.938 & 0.957 & 0.318 & 0.051 & 0.073 & 0.959 & 0.958 & 0.318 & 0.037 & 0.038 & 0.952 & 0.943 \\ 
  $\gamma_{z,4}$ & 0.316 & 0.318 & 0.071 & 0.088 & 0.957 & 0.960 & 0.316 & 0.053 & 0.061 & 0.948 & 0.939 & 0.314 & 0.036 & 0.038 & 0.950 & 0.942 \\ 
  $\sigma^2_{\eta_X}$ & 1.000 & 0.978 & 0.092 & 0.090 & 0.900 & 0.878 & 0.993 & 0.063 & 0.080 & 0.927 & 0.902 & 0.996 & 0.043 & 0.044 & 0.935 & 0.914 \\ 
  $\sigma^2_{\eta_Z}$ & 1.000 & 0.984 & 0.091 & 0.099 & 0.907 & 0.895 & 0.993 & 0.062 & 0.065 & 0.933 & 0.899 & 0.997 & 0.045 & 0.044 & 0.930 & 0.907 \\ 
  $\rho_{\eta_X \eta_Z}$ & 0.500 & 0.498 & 0.050 & 0.054 & 0.907 & 0.935 & 0.502 & 0.034 & 0.040 & 0.923 & 0.932 & 0.501 & 0.024 & 0.023 & 0.937 & 0.928 \\ 
   \hline

\hline 
\end{tabular}}
\caption{\textit{Simulation Results \textit{Setting 2}}}
\label{table:mc2}
\end{table}

Finally, we discuss some simulation evidence about the LR test in this setting. For both simulation schemes, we test the composite nulls that $\rho_U= \Z$ and $\rho_U = (0.5,0.5)^\prime$, respectively. When testing for $\rho_U= \Z$, the critical values are approximated by simulations, whereas for $\rho_U = (0.5,0.5)^\prime$, the critical values are obtained from a $\chi^2_2$. To obtain the critical values in the former case, we first numerically approximate of the score vector at $\hat\theta_{1,n}$ at each sample point. We then stack to it the closed-form expression of the (vectorized) Hessian matrix for $\rho_U$, which is relatively straightforward to estimate. Its expression is given in the proof of Proposition \ref{prop:identif2}. Finally, we compute the sample information matrix by taking the inner product of the augmented score matrix. Using the generalized inverse of the information matrix, we simulate $10000$ values from the distribution of $(Z_{\theta_1},Z_{\rho_U\rho_U^\prime})$, and we obtain an estimator of $\hat{\tau}_{\rho_U\rho_U^\prime}$ by a weighted projection of the realizations of $Z_{\rho_U\rho_U^\prime}$ into the positive orthant. This last step is performed through quadratic programming, as explained in \citet{andrews1999,andrews2001}.

Table \ref{table:sizetests} contains the size of the LR test. The nominal sizes are $\lbrace 10\%, 5\%,1\%\rbrace$, respectively. The columns indicate the true value of $\rho_U$ used in the simulation exercise and the null hypothesis of the test. For $\rho_U = \Z$, the test has size close to the nominal one, although it tends to be slightly conservative. When $\rho_U=(0.5,0.5)^\prime$, the LR test tends to have the opposite behavior as sizes are slightly larger than the nominal ones.

\begin{table}[!h]
\centering
\scriptsize
\begin{tabular}{m{40pt}|m{40pt}m{40pt}m{40pt}|m{40pt}m{40pt}m{40pt}}
\hline \hline
~ & \multicolumn{3}{c}{$\rho_U = \Z$, $H_0: \rho_U = \Z$} & \multicolumn{3}{c}{$\rho_U = (0.5,0.5)^\prime$, $H_0: \rho_U = (0.5,0.5)^\prime$} \\ \hline
% latex table generated in R 4.0.4 by xtable 1.8-4 package
% Thu Mar 18 00:31:52 2021
 & $250$ & $500$ & $1000$ & $250$ & $500$ & $1000$ \\ 
  \hline
0.1 & 0.002 & 0.075 & 0.075 & 0.131 & 0.111 & 0.115 \\ 
  0.05 & 0.001 & 0.047 & 0.037 & 0.069 & 0.067 & 0.066 \\ 
  0.01 & 0.001 & 0.020 & 0.008 & 0.014 & 0.017 & 0.015 \\ 
   \hline

\hline
\end{tabular}
\caption{\textit{Size of the Likelihood Ratio test}}
\label{table:sizetests}
\end{table}

In Table \ref{table:powertests}, we instead report the power properties of the LR test. The columns indicate the true value of $\rho_U$ used in the simulation exercise and the null hypothesis of the test. In general, the test has good power, and the power improves substantially as the sample size increases.

\begin{table}[!h]
\centering
\scriptsize
\begin{tabular}{m{40pt}|m{40pt}m{40pt}m{40pt}|m{40pt}m{40pt}m{40pt}}
\hline \hline
~ & \multicolumn{3}{c}{$\rho_U = \Z$, $H_0: \rho_U = (0.5,0.5)^\prime$} & \multicolumn{3}{c}{$\rho_U = (0.5,0.5)^\prime$, $H_0: \rho_U = \Z$} \\ \hline
% latex table generated in R 4.0.4 by xtable 1.8-4 package
% Thu Mar 18 00:31:52 2021
 & $250$ & $500$ & $1000$ & $250$ & $500$ & $1000$ \\ 
  \hline
0.1 & 0.000 & 0.940 & 0.991 & 0.937 & 0.998 & 1.000 \\ 
  0.05 & 0.000 & 0.917 & 0.990 & 0.888 & 0.995 & 1.000 \\ 
  0.01 & 0.000 & 0.839 & 0.989 & 0.763 & 0.986 & 1.000 \\ 
   \hline

\hline
\end{tabular}
\caption{\textit{Power of the Likelihood Ratio test}}
\label{table:powertests}
\end{table}

Finally, we report summary statistics for our estimators of technical efficiencies using the Battese-Coelli formula provided in equation \ref{eq:battesecoeli}. To give a reference point to the reader, in both simulation schemes the marginal distribution of $U$ is a half-normal distribution with scale parameter equal to $\sigma^2_U = 2.7519$. Therefore, the true mean technical efficiency is equal to 
\[
E\left[ \exp(-U) \right] = 2 \exp\left( \frac{\sigma^2_U}{2}\right) \Phi \left( -\sigma_U\right)  = 0.3846.
\]

Our estimator gives a plausible interval for the values of technical efficiencies. The mean technical efficiency also approaches its true value as the sample size increases. 

\begin{table}[!h]
\centering
\scriptsize
\begin{tabular}{l | c c | c c | c c} 
\hline \hline
~ & \multicolumn{2}{c}{$N = 250$} & \multicolumn{2}{c}{$N = 500$} & \multicolumn{2}{c}{$N = 1000$}\\ \hline
 & $\rho_U = \Z$ & $\rho_U = (0.5,0.5)^\prime$ & $\rho_U = \Z$ & $\rho_U = (0.5,0.5)^\prime$  & $\rho_U = \Z$ & $\rho_U = (0.5,0.5)^\prime$ \\ \hline
% latex table generated in R 4.0.4 by xtable 1.8-4 package
% Thu Mar 18 00:31:52 2021
  \hline
Min. & 0.000 & 0.000 & 0.000 & 0.000 & 0.000 & 0.000 \\ 
  1st Qu. & 0.227 & 0.244 & 0.232 & 0.241 & 0.230 & 0.238 \\ 
  Median & 0.419 & 0.430 & 0.408 & 0.415 & 0.399 & 0.408 \\ 
  Mean & 0.411 & 0.419 & 0.399 & 0.401 & 0.389 & 0.396 \\ 
  3rd Qu. & 0.581 & 0.586 & 0.560 & 0.561 & 0.548 & 0.554 \\ 
  Max. & 1.000 & 1.000 & 1.000 & 1.000 & 1.000 & 0.992 \\ 
   \hline

\hline
\end{tabular}
\caption{\textit{Summary measures for the estimator of technical efficiency}}
\label{table:techeff}
\end{table}

\section{Empirical Application} \label{sec:empiricalapp}

In this section, we consider an application using data on the agricultural sector in Nepal. The dataset consists of a cross-section of $600$ vegetable-cultivating farmers for the crop year 2015. The database is sourced from the International Food Policy Research Institute and the Seed Entrepreneurs' Association of Nepal (\citeyear{DataNepal}). For more detail on the data, see \cite{IFPRI2017}. The \emph{Output} variable is total vegetable production measured in rupees. \emph{Land} is measured as the total area cultivated in square feet. \emph{Labor} is the sum of hours worked by hired laborers and the hours worked by household members. \emph{Fertilizers} are the sum of organic and inorganic fertilizers, both measured in kilograms. \emph{Seeds} are measured as the sum of hybrid and pollinated seeds in grams. As environmental variables we consider \emph{Education}, as the proportion of household members with higher education or professional degree; \emph{Experience}, which is the number of years the farmer has been growing vegetables; and an indicator of risk diversification, \emph{Risk Div}, which is constructed as an Ogive index of relative economic diversification \citep{wasylenko1978}. It is defined as
\begin{equation*}
Risk\;Div\;= \sum_{i=1}^{N_C}\frac{(s_{i} - \bar{s}_i)^{2}}{\bar{s}_i},
\end{equation*}
where $s_{i}$ is the proportion of land devoted by the farmer to crop $i$, $\bar{s}_i$ is the average sample proportion of land devoted to crop $i$, and $N_C$ is the total number of crops cultivated by each farmer. A higher value of the \emph{Risk Div} index implies lower risk diversification. After removing missing values, we obtain a final sample of $497$ observations. Summary statistics of the variables used in the analysis are provided in Appendix \ref{sec:appC}.

The model we estimate is the following
\[
Y = X\beta + V - U_0 \exp(Z\delta),
\]
where
\begin{align*}
Y =& \lbrace \log(Output) \rbrace,\\
X =& \lbrace Intercept,\log(Land), \log(Labor),\log(Fertilizers),\log(Seeds) \rbrace,\\
Z =& \lbrace Education, Experience, \log(Risk\; Div) \rbrace.
\end{align*}

We allow for endogeneity of three inputs (\emph{Labor}, \emph{Fertilizers}, and \emph{Seeds}) and one environmental variable (\emph{Risk Div}). As instruments, we use a dummy for whether the farmer has suffered any natural shocks in the two years prior to the survey (\emph{Natural Shocks}); the average years of experience of nearby farmers, as a measure of spillover effects (\emph{Peers Experience}), and its square; three variables measuring the proportion of seeds that are owned by the farmer (\emph{Own Supplier}), obtained through formal channels such as an input retailer, a private seed company or representative, a government extension service or a research institute (\emph{Formal Supplier}), or informal channels such as a family member, a farmer's cooperative, gifted from a nearby farmer, friend or farmer from other villages, or landlord (\emph{Informal Supplier}); and interaction terms between these variables. \footnote{We have checked for weak instruments using the Cragg–Donald statistic, $CG_n$ \citep[see][]{cragg1993,stock2005}. We obtain a value of $CG_n = 4.891$. Our instruments appear to be sufficiently strong based on the critical values reported in Table 5.4 of \citet{stock2005}, with $2$ endogenous variables, $10$ instruments and a maximum size distortion between $10\%$ and $15\%$. This conclusion is speculative, as we do not know what the distribution of our estimator is under weak-instrument asymptotic. We have also computed the value of the Cragg–Donald statistic in our simulation study with two instruments and two endogenous variables and $N = 500$. The $95$ percentile of the test statistic is equal to $1.919$, which seems to support our conclusion.}

In this application, we normalize the dependence parameter between \emph{Fertilizers} and $U_0$, $\rho_{U,\eta_{Fertilizer}}$, to be positive. The motivation for this choice is that the use of \emph{Fertilizers} for production may be related to soil quality, unobserved by the econometrician and which ultimately influences the efficiency of the producer. We have run some robustness checks and our results are not sensitive to the choice of normalization.

\begin{table}[!h]
\centering
\scriptsize
\begin{tabular}{l | c l r | c l r | c l r} 
\hline \hline
~ & \multicolumn{3}{c}{Exogeneity} & \multicolumn{3}{c}{Endogeneity, $\rho_U=\Z$} & \multicolumn{3}{c}{Endogeneity} \\ \hline
~ & Estimate & \multicolumn{2}{c}{95\% CI}  & Estimate & \multicolumn{2}{c}{95\% CI} & Estimate & \multicolumn{2}{c}{95\% CI} \\ \hline
% latex table generated in R 4.0.4 by xtable 1.8-4 package
% Thu Mar 18 00:32:12 2021
  \hline
$\beta_0$ & 8.0440 & [4.0438 & 9.4146] & 6.2405 & [-0.0110 & 8.9503] & 6.4094 & [4.3764 & 8.3598] \\ 
  $\beta_{Land}$ & 0.1721 & [0.0606 & 0.4524] & -0.1116 & [-0.3935 & 0.4632] & -0.0988 & [-0.2155 & 0.1780] \\ 
  $\beta_{Labor}$ & 0.1423 & [0.0056 & 0.2529] & 0.2593 & [-0.5882 & 0.8927] & 0.2966 & [0.0149 & 0.5558] \\ 
  $\beta_{Fertilizer}$ & 0.0940 & [0.0387 & 0.3141] & 0.4201 & [0.1015 & 0.9596] & 0.3914 & [0.1453 & 0.5640] \\ 
  $\beta_{Seeds}$ & 0.1171 & [0.0323 & 0.2388] & 0.4913 & [0.1785 & 0.8723] & 0.4608 & [0.2202 & 0.6008] \\ 
  $\delta_{Education}$ & 0.1726 & [-34.3189 & 1.3928] & -0.9533 & [-8.6051 & 5.6725] & -0.7039 & [-2.3566 & 1.0696] \\ 
  $\delta_{Experience}$ & -0.2986 & [-4.7863 & 5.2293] & 2.8778 & [-1.2931 & 6.3602] & 2.2386 & [0.1417 & 3.0539] \\ 
  $\delta_{Risk}$ & -0.4340 & [-1.1017 & 4.0653] & -0.1069 & [-0.9274 & 0.8136] & -0.0988 & [-0.2125 & 0.3224] \\ 
  $\rho_{U,\eta_{Labor}}$ & ~ & ~ & ~ & ~ & ~ & ~ & -0.1375 & [-0.5201 & 0.4231] \\ 
  $\rho_{U,\eta_{Fertilizer}}$ & ~ & ~ & ~ & ~ & ~ & ~ & 0.4493 & [0.1963 & 0.6873] \\ 
  $\rho_{U,\eta_{Seeds}}$ & ~ & ~ & ~ & ~ & ~ & ~ & -0.3383 & [-0.4938 & 0.3606] \\ 
  $\rho_{U,\eta_{Risk}}$ & ~ & ~ & ~ & ~ & ~ & ~ & 0.1604 & [-0.4605 & 0.4941] \\ 
  $\rho_{V,\eta_{Labor}}$ & ~ & ~ & ~ & -0.3045 & [-0.5124 & 0.0655] & -0.3350 & [-0.5150 & -0.0594] \\ 
  $\rho_{V,\eta_{Fertilizer}}$ & ~ & ~ & ~ & -0.4445 & [-0.5898 & -0.0500] & -0.4127 & [-0.5703 & 0.0035] \\ 
  $\rho_{V,\eta_{Seeds}}$ & ~ & ~ & ~ & -0.5282 & [-0.6572 & -0.2132] & -0.5272 & [-0.6392 & -0.2661] \\ 
  $\rho_{V,\eta_{Risk}}$ & ~ & ~ & ~ & 0.1121 & [-0.0619 & 0.2489] & 0.1140 & [-0.0335 & 0.2668] \\ 
  $\sigma^2_{U}$ & 0.0100 & [0.0069 & 1.5672] & 0.0347 & [0.0237 & 1.5798] & 0.0811 & [0.0554 & 1.6315] \\ 
  $\sigma^2_{V}$ & 1.0105 & [0.6900 & 4.3150] & 1.6148 & [1.1026 & 4.7276] & 1.4830 & [1.0126 & 2.0437] \\ 
   \hline

\hline \hline 
\end{tabular}
\caption{\textit{Estimates of the production function parameters for farmers in Nepal.}}
\label{table:nepalres}
\end{table}

Table \ref{table:nepalres} reports the estimated coefficients and the $95\%$ confidence intervals for our empirical example. The confidence intervals are obtained using $496$ subsamples of size $b = 50$.

The left panel shows the estimation results assuming exogeneity. All the estimated coefficients for inputs are positive and significant, although generally small in magnitude, being \emph{Labor} and \emph{Land}, the inputs with the largest estimated effect. None of the environmental variables appears to have a significant effect on inefficiency. The exogenous model detects very little inefficiency, and the parameter $\delta$ is estimated imprecisely, as it can be noticed by the length of the confidence intervals.

In the center panel, we report the estimation results controlling for endogeneity but restricting $\rho_U = \Z$, i.e., imposing independence between the endogenous variables and the inefficiency term. As it is usually the case in instrumental variable models, confidence intervals are wider than in the model assuming exogeneity. However, controlling for endogeneity substantially changes the conclusions obtained from this empirical example, regarding the effect of inputs, but especially the effect of the inefficiency determinants. We find that most of the estimated coefficients for the inputs are positive but significant only for \emph{Fertilizers} and \emph{Seeds}. Regarding the environmental variables, the estimated coefficient of \emph{Risk Div} is negative, which means that farmers cultivating fewer crops (i.e., with lower risk diversification) are more efficient, possibly due to higher specialization levels. However, the effect is not significant at the $5\%$ level. The correlation between the two-sided error term and the endogenous variables is negative and significant, except for \emph{Risk Div}.

The right panel shows the results controlling for endogeneity without restricting the dependence between the endogenous variables and the inefficiency term. The estimated coefficients are quite similar to those when we impose that $\rho_U = 0$, except for \emph{Experience} whose coefficient is positive and significant in the third model. This result may be due to more experienced, i.e. older, farmers being more conservative, and therefore, less willing to implement new practices that could improve their efficiency \citep{coelli1996}. The $95\%$ confidence intervals are much narrower than those for the restricted estimator and indicate that only the choice of \emph{Fertilizers} may be related to the producer's inefficiency level. We test for the absence of dependence between the endogenous variables and the inefficiency term being equal to $0$ using the LR test, as explained in the simulation study. The value of the test statistic is equal to $6.77$, and we obtain a $95\%$ critical value equal to $11.41$. Hence, we cannot reject the null that $\rho_U = \Z$. 

We also test the null that $\sigma^2_U=0$ in the three models. Under the null, $\rho_U$ and $\delta$ are nuisance parameters. As we do not know the asymptotic distribution of the LR test statistics in this case and we are testing a parameter at the boundary, critical values are obtained from an equal mixture of a mass point at $0$ and a $\chi^2$-distribution with $1$ degree of freedom \citep{lee1993,andrews2001,ketz2018}. In both the exogenous and the endogenous model, we reject the null of no inefficiency at the $1\%$ level. The value of the LR statistic is $4.66$ for the exogenous model, and $79.46$ for the endogenous model, with a critical value equal to $3.82$.

\begin{figure}[!h]
\includegraphics[scale=0.6]{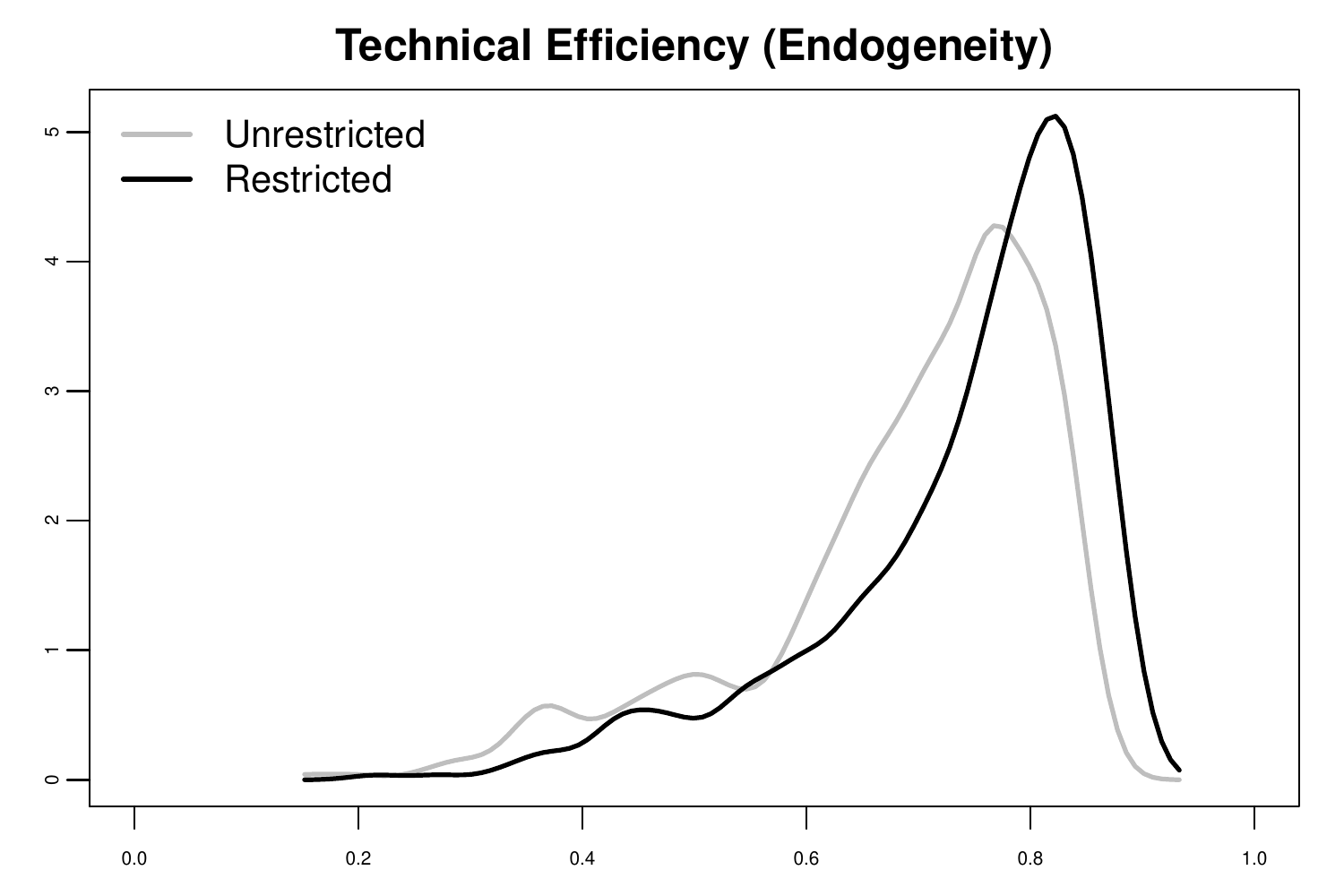}
\caption{Estimation of technical efficiency.}
\label{fig:fct_nepal_eff}
\end{figure}

Finally, Figure \ref{fig:fct_nepal_eff} reports the kernel density estimator of technical efficiency for the endogenous models. The density of the efficiency scores is similar in both models, which is an expected result given that we cannot reject the null that $\rho_U$ is equal to 0.

\section{Conclusions} \label{sec:conclusion}

We propose and study an estimator of stochastic frontier models when both the production inputs and the environmental variables are correlated with the two-sided stochastic error term and the one-sided stochastic inefficiency term. Our identification and estimation strategy is based on control functions that fully capture the dependence between regressors and unobservables. While the joint density of the two-sided stochastic error term and the control function is modeled as a normal distribution, one of the main challenges for direct maximum likelihood estimation is to write the joint density of the stochastic inefficiency term and the control function in closed form. To circumvent this issue, \citet{amsler2017} use copula functions to model the dependence between observables and unobservables components of the model, and employ a simulated maximum likelihood procedure to obtain the parameter's estimate. This estimator may not be easy to implement and may be computationally slow. Moreover, instrumental variable methods lead to lower precision in the estimate and simulated methods can increase this lack of precision even further.

In this work, we provide a simple maximum likelihood estimator that aims at avoiding these potential pitfalls. Our main assumption is that the conditional distribution of the stochastic inefficiency term given the control functions is a folded normal distribution. This distribution reduces to the half-normal when there is no endogeneity. This makes our model a straightforward extension of the normal-half-normal model to include endogenous regressors. We shed light on some new identification issues, and we provide some theoretical results on estimation and inference. Our estimator is easy and fast to implement, and enjoys good finite sample properties.

Additional research on the properties of the trinity of tests and on testing the distributional assumptions on the error term is needed. Moreover, extensions of our model to panel data with time-varying endogeneity and true fixed effects could be of interest.

%%%%%%%%%%%%%%%%%%%%%%%%%%%%%%%%%%%%%%%%%%%%%%%%%%%%%%%%%%%%%%%%%%%%%%%%%%%%%%%%
%%%%%%%%%%%%%%%%%%%%%%%%%%%%%%%%% REFERENCES %%%%%%%%%%%%%%%%%%%%%%%%%%%%%%%%%%%
%%%%%%%%%%%%%%%%%%%%%%%%%%%%%%%%%%%%%%%%%%%%%%%%%%%%%%%%%%%%%%%%%%%%%%%%%%%%%%%%
\bibliography{efficiency}
\bibliographystyle{agsm}

% \clearpage % ensure all floats are processed
% \processdelayedfloats
% \clearpage

\newpage

\appendix
\renewcommand{\theequation}{\thesection.\arabic{equation}}
\setcounter{equation}{0}

\section{Main Proofs} \label{sec:appA}

\subsection{Marginal density of $U_0$}

We start from the conditional density of $U_0$ given $\eta$, and the marginal density of $\eta$ given by
\begin{align*}
f_{U_0 \vert \eta}(u \vert \eta) =& \frac{1}{\sqrt{2\pi (\sigma_U^2 - \Sigma^\prime_{U\eta}\Sigma^{-1}_\eta \Sigma_{U\eta})}}\left\lbrace \exp \left( -\frac{(u - \Sigma^\prime_{U\eta} \Sigma^{-1}_\eta\eta)^2}{2(\sigma_U^2 - \Sigma^\prime_{U\eta}\Sigma^{-1}_\eta \Sigma_{U\eta})}\right) + \exp \left( -\frac{(u + \Sigma^\prime_{U\eta} \Sigma^{-1}_\eta\eta)^2}{2(\sigma_U^2 - \Sigma^\prime_{U\eta}\Sigma^{-1}_\eta \Sigma_{U\eta})}\right)\right\rbrace,\\
f_{\eta}(\eta) =& \frac{1}{(2\pi)^{\frac{p+k}{2}}}\vert \Sigma_\eta \vert^{-\frac{1}{2}} \exp \left( -\frac{1}{2} \eta^\prime \Sigma^{-1}_\eta \eta \right),
\end{align*}
where $\vert \Sigma_\eta \vert$ denotes the determinant of $\Sigma_\eta$, and we let $\Sigma_{U\eta} = D_\eta \rho_U \sigma_U$, to simplify notations.

We wish to prove that the marginal density of $U_0$ is the half-normal density. First notice that, by block matrix inversion, and the matrix inversion lemma \citep{sherman1950}, we have that
\begin{align*}
\begin{bmatrix} \sigma_U^2 & \pm \Sigma^\prime_{U\eta} \\ \pm \Sigma_{U\eta} & \Sigma_{\eta}\end{bmatrix}^{-1} =& \begin{bmatrix} \left( \sigma_U^2 - \Sigma^\prime_{U\eta} \Sigma^{-1}_\eta \Sigma_{U\eta}\right)^{-1} & \mp\left( \sigma_U^2 - \Sigma^\prime_{U\eta} \Sigma^{-1}_\eta \Sigma_{U\eta}\right)^{-1} \Sigma^\prime_{U\eta}\Sigma_\eta^{-1} \\ \mp\Sigma_\eta^{-1} \Sigma_{U\eta} \left( \sigma_U^2 - \Sigma^\prime_{U\eta} \Sigma^{-1}_\eta \Sigma_{U\eta}\right)^{-1} & \Sigma^{-1}_{\eta} + \Sigma^{-1}_{\eta} \Sigma_{U\eta}\left( \sigma_U^2 - \Sigma^\prime_{U\eta} \Sigma^{-1}_\eta \Sigma_{U\eta}\right)^{-1} \Sigma^\prime_{U\eta} \Sigma^{-1}_\eta \end{bmatrix}\\
=& \begin{bmatrix} \frac{1}{\sigma_U^2} + \frac{\Sigma^\prime_{U\eta}}{\sigma_U^2} \left( \Sigma_\eta  - \frac{\Sigma_{U\eta}\Sigma^\prime_{U\eta}}{\sigma^2_U}\right)^{-1} \frac{\Sigma_{U\eta}}{\sigma_U^2} & \mp\frac{\Sigma^\prime_{U\eta}}{\sigma_U^2} \left( \Sigma_\eta  - \frac{\Sigma_{U\eta}\Sigma^\prime_{U\eta}}{\sigma^2_U}\right)^{-1} \\ \mp\left( \Sigma_\eta  - \frac{\Sigma_{U\eta}\Sigma^\prime_{U\eta}}{\sigma^2_U}\right)^{-1}\frac{\Sigma_{U\eta}}{\sigma_U^2}  & \left( \Sigma_\eta  - \frac{\Sigma_{U\eta}\Sigma^\prime_{U\eta}}{\sigma^2_U}\right)^{-1} \end{bmatrix}.
\end{align*}
Let
\[
\Omega_+ = \begin{bmatrix} \sigma_U^2 & \Sigma^\prime_{U\eta} \\ \Sigma_{U\eta} & \Sigma_{\eta}\end{bmatrix}, \text{ and } \Omega_{-} = \begin{bmatrix} \sigma_U^2 & -\Sigma^\prime_{U\eta} \\ -\Sigma_{U\eta} & \Sigma_{\eta}\end{bmatrix}
\]

Using results about determinants of block matrices, we have that 
\[
\vert \Omega_+ \vert = \vert \Omega_{-} \vert = \left( \sigma_U^2 - \Sigma^\prime_{U\eta} \Sigma^{-1}_\eta \Sigma_{U\eta}\right) \vert \Sigma_\eta \vert = \sigma_U^{2} \left\vert \Sigma_\eta  - \frac{\Sigma_{U\eta}\Sigma^\prime_{U\eta}}{\sigma^2_U} \right\vert. 
\]
Therefore, using a similar decomposition of the conditional density as above, we obtain
\begin{align*}
f_{U_0, \eta,1} & (u , \eta)= f_{U_0 \vert \eta,1}(u \vert \eta)f_\eta(\eta) \\
=& \frac{\vert \Sigma_\eta \vert^{-\frac{1}{2}}}{\sqrt{(2\pi)^{p+k+1} (\sigma_U^2 - \Sigma^\prime_{U\eta}\Sigma^{-1}_\eta \Sigma_{U\eta})}} \exp \left( -\frac{(u - \Sigma^\prime_{U\eta} \Sigma^{-1}_\eta\eta)^2}{2(\sigma_U^2 - \Sigma^\prime_{U\eta}\Sigma^{-1}_\eta \Sigma_{U\eta})} - \frac{1}{2} \eta^\prime \Sigma^{-1}_\eta \eta \right) \\
=& \frac{1}{(2\pi)^{\frac{p+k+1}{2}}}\vert \Omega_+ \vert^{-\frac{1}{2}} \exp \left( -\frac{1}{2}\begin{pmatrix} u \\ \eta \end{pmatrix}^\prime \Omega^{-1}_+ \begin{pmatrix} u \\ \eta \end{pmatrix} \right) \\
=& \frac{1}{(2\pi)^{\frac{p+k}{2}}}\left\vert \Sigma_\eta  - \frac{\Sigma_{U\eta}\Sigma^\prime_{U\eta}}{\sigma^2_U} \right\vert^{-\frac{1}{2}} \exp \left( -\frac{1}{2}\left( \eta - \frac{\Sigma_{U\eta}u}{\sigma^2_U} \right)^\prime \left(  \Sigma_\eta  - \frac{\Sigma_{U\eta}\Sigma^\prime_{U\eta}}{\sigma^2_U} \right)^{-1} \left( \eta - \frac{\Sigma_{U\eta}u}{\sigma^2_U} \right) \right) \times \\
& \qquad \frac{1}{\sqrt{2\pi \sigma^2_U}} \exp(-\frac{u^2}{2\sigma^2_U}),
\end{align*}
where the first term is the density of a normal distribution with mean $\Sigma_{U\eta}u/\sigma^2_U$ and variance $\Sigma_\eta  - \Sigma_{U\eta}\Sigma^\prime_{U\eta}/\sigma^2_U$, and whose integral with respect to $\eta \in (-\infty,\infty)$ is therefore equal to one. Similarly, we have
\begin{align*}
f_{U_0, \eta,2}& (u , \eta)= f_{U_0 \vert \eta,2}(u \vert \eta)f_\eta(\eta) \\
=& \frac{\vert \Sigma_\eta \vert^{-\frac{1}{2}}}{\sqrt{(2\pi)^{p+k+1} (\sigma_U^2 + \Sigma^\prime_{U\eta}\Sigma^{-1}_\eta \Sigma_{U\eta})}} \exp \left( -\frac{(u + \Sigma^\prime_{U\eta} \Sigma^{-1}_\eta\eta)^2}{2(\sigma_U^2 - \Sigma^\prime_{U\eta}\Sigma^{-1}_\eta \Sigma_{U\eta})} - \frac{1}{2} \eta^\prime \Sigma^{-1}_\eta \eta \right) \\
=& \frac{1}{(2\pi)^{\frac{p+k+1}{2}}}\vert \Omega_{-} \vert^{-\frac{1}{2}} \exp \left( -\frac{1}{2}\begin{pmatrix} u \\ \eta \end{pmatrix}^\prime \Omega^{-1}_{-} \begin{pmatrix} u \\ \eta \end{pmatrix} \right) \\
=& \frac{1}{(2\pi)^{\frac{p+k}{2}}}\left\vert \Sigma_\eta - \frac{\Sigma_{U\eta}\Sigma^\prime_{U\eta}}{\sigma^2_U} \right\vert^{-\frac{1}{2}} \exp \left( -\frac{1}{2}\left( \eta + \frac{\Sigma_{U\eta}u}{\sigma^2_U} \right)^\prime \left(  \Sigma_\eta  - \frac{\Sigma_{U\eta}\Sigma^\prime_{U\eta}}{\sigma^2_U} \right)^{-1} \left( \eta + \frac{\Sigma_{U\eta}u}{\sigma^2_U} \right) \right) \times \\
& \qquad \frac{1}{\sqrt{2\pi \sigma^2_U}} \exp(-\frac{u^2}{2\sigma^2_U}),
\end{align*}
where the first term is the density of a normal distribution with mean $-\Sigma_{U\eta}u/\sigma^2_U$ and variance $\Sigma_\eta  - \Sigma_{U\eta}\Sigma^\prime_{U\eta}/\sigma^2_U$, and whose integral with respect to $\eta \in (-\infty,\infty)$ is therefore equal to one. 

Hence, 
\begin{align*}
f_{U_0}(u ) = \int_{-\infty}^{\infty} f_{U_0, \eta}(u , \eta) d\eta =&  \int_{-\infty}^{\infty} f_{U_0, \eta,2}(u , \eta) d\eta + \int_{-\infty}^{\infty} f_{U_0, \eta,2}(u , \eta) d\eta \\
=& \frac{2}{\sqrt{2\pi \sigma^2_U}} \exp \left( -\frac{u^2}{2\sigma^2_U}\right),
\end{align*}
which is the density of a half-normal distribution. This concludes the proof.

\subsection{Conditional density of the composite error term}

In this subsection, we provide the main steps to derive the conditional density of the composite error term, $\varepsilon$, given $\eta$. As above, we let $\Sigma_{V\eta} =  D_\eta \rho_V \sigma_V$, and $\Sigma_{U\eta} =  D_\eta \rho_U \sigma_U$, to simplify notations. Recall that 
\begin{align*}
f_{V\vert\eta} & (\varepsilon + u \vert \eta ) \left( g(Z,\delta )\right)^{-1} f_{U_0\vert \eta }( \left( g(Z,\delta )\right)^{-1} u \vert \eta)  \\
=& \frac{1}{2\pi \tilde{\sigma}_U (Z) \tilde{\sigma}_V} \left\lbrace \exp \left( -\frac{(u - g \left( Z,\delta\right) \Sigma_{U\eta}^\prime \Sigma^{-1}_\eta\eta)^2}{2\tilde{\sigma}^2_U(Z)}- \frac{(\varepsilon + u - \Sigma_{V\eta}^\prime\Sigma^{-1}_\eta\eta)^2}{2\tilde{\sigma}^2_V}\right) \right. \\
&\quad \left. + \exp \left( -\frac{(u + g \left( Z,\delta\right) \Sigma_{U\eta}^\prime \Sigma^{-1}_\eta\eta)^2}{2\tilde{\sigma}_U(Z)^2}- \frac{(\varepsilon + u - \Sigma_{V\eta}^\prime\Sigma^{-1}_\eta\eta)^2}{2\tilde{\sigma}^2_V}\right) \right\rbrace,
\end{align*}
where $\tilde{\sigma}^2_U(Z) = (\sigma^2_U -  \Sigma_{U\eta}^\prime\Sigma^{-1}_\eta \Sigma_{U\eta})g^2 \left( Z,\delta\right)$, and $\tilde{\sigma}^2_V = \sigma^2_V - \Sigma_{V\eta}^\prime\Sigma^{-1}_\eta \Sigma_{V\eta}$. 

The terms inside the exponential function can be treated similarly, and for simplicity, we only show the algebra for the first term. We have 
\begin{align*}
\frac{(u - g \left( Z,\delta\right) \Sigma_{U\eta}^\prime \Sigma^{-1}_\eta\eta)^2}{\tilde{\sigma}_U(Z)} =& \frac{1}{\tilde{\sigma}^2_U(Z)} \left( u^2 - 2 g \left( Z,\delta\right) \Sigma_{U\eta}^\prime \Sigma^{-1}_\eta \eta u + \left( g \left( Z,\delta\right) \Sigma_{U\eta}^\prime \Sigma^{-1}_\eta \eta\right)^2 \right)\\
\frac{(\varepsilon + u - \Sigma_{V\eta}^\prime\Sigma^{-1}_\eta\eta)^2}{\tilde{\sigma}^2_V} =& \frac{1}{\tilde{\sigma}^2_V} \left( u^2 + \left( \varepsilon - \Sigma_{V\eta}^\prime\Sigma^{-1}_\eta \eta \right)^2 + 2\left( \varepsilon - \Sigma_{V\eta}^\prime\Sigma^{-1}_\eta \eta \right) u \right).
\end{align*}

Taking the sum of these two terms gives 
\begin{align*}
\frac{\sigma^2(Z)}{\tilde{\sigma}^2_U(Z) \tilde{\sigma}^2_V} & \left( u^2 - 2 g \left( Z,\delta\right) \Sigma_{U\eta}^\prime \Sigma^{-1}_\eta \eta u  \frac{\tilde{\sigma}^2_V}{\sigma^2(Z)} + 2\left( \varepsilon - \Sigma_{V\eta}^\prime\Sigma^{-1}_\eta \eta \right) u \frac{\tsigma^2_U(Z)}{\sigma^2(Z)}\right) \\
&\quad + \frac{\left( g \left( Z,\delta\right) \Sigma_{U\eta}^\prime \Sigma^{-1}_\eta\eta \right)^2 }{\tilde{\sigma}^2_U(Z)}  + \frac{\left( \varepsilon - \Sigma_{V\eta}^\prime\Sigma^{-1}_\eta \eta \right)^2}{\tilde{\sigma}^2_V} \\
=& \frac{\sigma^2(Z)}{\tilde{\sigma}^2_U(Z) \tilde{\sigma}^2_V} \left( u + \left( \varepsilon - \Sigma_{V\eta}^\prime\Sigma^{-1}_\eta \eta \right)\frac{\tilde{\sigma}^2_U(Z)}{\sigma^2(Z)} - g \left( Z,\delta\right) \Sigma_{U\eta}^\prime \Sigma^{-1}_\eta \eta  \frac{\tilde{\sigma}^2_V}{\sigma^2(Z)}\right)^2 \\
&\quad -\frac{\sigma^2(Z)}{\tilde{\sigma}^2_U(Z) \tilde{\sigma}^2_V}\left( \left( \varepsilon - \Sigma_{V\eta}^\prime\Sigma^{-1}_\eta \eta \right)\frac{\tilde{\sigma}^2_U(Z)}{\sigma^2(Z)} - g \left( Z,\delta\right) \Sigma_{U\eta}^\prime \Sigma^{-1}_\eta \eta  \frac{\tilde{\sigma}^2_V}{\sigma^2(Z)}\right)^2 \\
&\quad + \frac{\left( g \left( Z,\delta\right) \Sigma_{U\eta}^\prime \Sigma^{-1}_\eta  \eta\right)^2 }{\tilde{\sigma}^2_U(Z)}  + \frac{\left( \varepsilon - \Sigma_{V\eta}^\prime\Sigma^{-1}_\eta \eta \right)^2}{\tilde{\sigma}^2_V} \\
=& \frac{\sigma^2(Z)}{\tilde{\sigma}^2_U(Z) \tilde{\sigma}^2_V} \left[ u + \left( \varepsilon - \Sigma_{V\eta}^\prime\Sigma^{-1}_\eta \eta \right)\frac{\tilde{\sigma}^2_U(Z)}{\sigma^2(Z)} - g \left( Z,\delta\right) \Sigma_{U\eta}^\prime \Sigma^{-1}_\eta \eta  \frac{\tilde{\sigma}^2_V}{\sigma^2(Z)}\right]^2 \\
&\quad + \left( \frac{1}{\tilde{\sigma}^2_V}  - \frac{\tilde{\sigma}^2_U(Z)}{\tilde{\sigma}^2_V \sigma^2(Z)} \right)\left( \varepsilon - \Sigma_{V\eta}^\prime\Sigma^{-1}_\eta \eta \right)^2 + \left( \frac{1}{\tilde{\sigma}^2_U(Z)}  - \frac{\tilde{\sigma}^2_V}{\tilde{\sigma}^2_U(Z) \sigma^2(Z)} \right) \left( g \left( Z,\delta\right) \Sigma_{U\eta}^\prime \Sigma^{-1}_\eta \eta \right)^2 \\
&\quad + \frac{2}{\sigma^2(Z)} \left( \varepsilon - \Sigma_{V\eta}^\prime\Sigma^{-1}_\eta \eta \right) g \left( Z,\delta\right) \Sigma_{U\eta}^\prime \Sigma^{-1}_\eta \eta \\
=& \frac{\sigma^2(Z)}{\tilde{\sigma}^2_U(Z) \tilde{\sigma}^2_V} \left[ u + \left( \varepsilon - \Sigma_{V\eta}^\prime\Sigma^{-1}_\eta \eta \right)\frac{\tilde{\sigma}^2_U(Z)}{\sigma^2(Z)} - g \left( Z,\delta\right) \Sigma_{U\eta}^\prime \Sigma^{-1}_\eta \eta  \frac{\tilde{\sigma}^2_V}{\sigma^2(Z)}\right]^2 \\
& \quad +\frac{1}{\sigma^2(Z)} \left( \varepsilon - \Sigma_{V\eta}^\prime\Sigma^{-1}_\eta \eta + g \left( Z,\delta\right) \Sigma_{U\eta}^\prime \Sigma^{-1}_\eta \eta \right)^2.
\end{align*}

Then, treating the remaining term similarly, we can write 
\begin{align*}
f_{V\vert\eta} & (\varepsilon + u \vert \eta ) \left( g(Z,\delta )\right)^{-1} f_{U_0\vert \eta }( \left( g(Z,\delta )\right)^{-1} u \vert \eta)  \\
=& \frac{1}{2\pi \frac{\tilde{\sigma}_U (Z) \tilde{\sigma}_V}{\sigma(Z)} \sigma(Z) } \left\lbrace \exp \left( -\frac{  \sigma^2(Z) \left[ u + \left( \left( \varepsilon - \Sigma_{V\eta}^\prime\Sigma^{-1}_\eta \eta \right)\frac{\tilde{\sigma}^2_U(Z)}{\sigma^2(Z)} - g \left( Z,\delta\right) \Sigma_{U\eta}^\prime \Sigma^{-1}_\eta \eta  \frac{\tilde{\sigma}^2_V}{\sigma^2(Z)}\right)\right]^2}{2\tilde{\sigma}^2_U(Z)\tilde{\sigma}^2_V} \right) \times  \right.\\
& \quad \exp \left(- \frac{\left(\varepsilon - \Sigma_{V\eta}^\prime\Sigma^{-1}_\eta \eta + g \left( Z,\delta\right) \Sigma_{U\eta}^\prime \Sigma^{-1}_\eta \eta\right)^2}{2\sigma^2(Z) }\right)  \\
&\quad + \exp \left( -\frac{  \sigma^2(Z) \left[ u + \left( \left( \varepsilon - \Sigma_{V\eta}^\prime\Sigma^{-1}_\eta \eta \right)\frac{\tilde{\sigma}^2_U(Z)}{\sigma^2(Z)} + g \left( Z,\delta\right) \Sigma_{U\eta}^\prime \Sigma^{-1}_\eta \eta  \frac{\tilde{\sigma}^2_V}{\sigma^2(Z)}\right)\right]^2}{2\tilde{\sigma}^2_U(Z)\tilde{\sigma}^2_V} \right) \times \\
& \quad \left. \exp \left(- \frac{\left(\varepsilon - \Sigma_{V\eta}^\prime\Sigma^{-1}_\eta \eta - g \left( Z,\delta\right) \Sigma_{U\eta}^\prime \Sigma^{-1}_\eta \eta\right)^2}{2\sigma^2(Z) }\right) \right\rbrace.
\end{align*}

By integrating the last expression with respect to $u \in (0,\infty)$, we obtain,
\begin{align*}
f_{\varepsilon\vert\eta} & (\varepsilon \vert \eta ) = \frac{1}{\sqrt{2\pi} \sigma(Z)} \left\lbrace \Phi \left( \frac{g\left( Z,\delta\right) \Sigma_{U\eta}^\prime \Sigma^{-1}_\eta \eta}{\lambda(Z) \sigma(Z)} -\frac{\lambda(Z) ( \varepsilon - \Sigma_{V\eta}^\prime \Sigma^{-1}_\eta \eta)}{\sigma(Z)}\right) \times \right. \\
& \quad \exp \left( -\frac{(\varepsilon -  \Sigma_{V\eta}^\prime \Sigma^{-1}_\eta \eta + g\left( Z,\delta\right)  \Sigma_{U\eta}^\prime \Sigma^{-1}_\eta \eta)^2}{2 \sigma^2(Z)}\right)\\
& \quad + \Phi \left( - \frac{g \left( Z,\delta\right)  \Sigma_{U\eta}^\prime \Sigma^{-1}_\eta \eta}{\lambda(Z)\sigma(Z)} -\frac{\lambda(Z) ( \varepsilon - \Sigma_{V\eta}^\prime \Sigma^{-1}_\eta \eta)}{\sigma(Z)}\right) \times \\
& \quad \left. \exp \left( - \frac{(\varepsilon -\Sigma_{V\eta}^\prime \Sigma^{-1}_\eta \eta - g \left( Z,\delta\right) \Sigma_{U\eta}^\prime \Sigma^{-1}_\eta \eta)^2}{2 \sigma^2(Z)}\right) \right\rbrace.
\end{align*}
Finally, the normalizing constants are computed using Lemma 2.2. in \citet[][p. 26]{azzalini2013}, which implies that
\begin{align*}
& \int_{-\infty}^{+\infty} \frac{1}{\sqrt{2\pi} \sigma(Z)} \Phi \left( \pm \frac{g\left( Z,\delta\right) \Sigma_{U\eta}^\prime \Sigma^{-1}_\eta \eta}{\lambda(Z) \sigma(Z)} -\frac{\lambda(Z) ( \varepsilon - \Sigma_{V\eta}^\prime \Sigma^{-1}_\eta \eta)}{\sigma(Z)}\right) \exp \left( -\frac{(\varepsilon -  \Sigma_{V\eta}^\prime \Sigma^{-1}_\eta \eta + g\left( Z,\delta\right)  \Sigma_{U\eta}^\prime \Sigma^{-1}_\eta \eta)^2}{2 \sigma^2(Z)}\right) d\varepsilon \\
=& \Phi \left( \pm \frac{g\left( Z,\delta\right) \Sigma_{U\eta}^\prime \Sigma^{-1}_\eta \eta}{\lambda(Z) \sigma(Z)}\frac{1}{\sqrt{1 + \lambda^2(Z)}}\right) = \Phi \left( \pm \frac{\tsigma^2_V g\left( Z,\delta\right) \Sigma_{U\eta}^\prime \Sigma^{-1}_\eta \eta}{\tsigma_U(Z) \sigma^2(Z)}\right).
\end{align*}
This concludes the proof.

\subsection{Proof of Proposition \ref{prop:identif1}}

Let $R = (X^\prime,Z^\prime)^\prime$, with
\[
R  = W \gamma +\eta. 
\]
The conditional density of $\varepsilon$ given $\eta$ is
\begin{equation}\label{eq:conddens1}
\begin{aligned}
f_{\varepsilon \vert \eta} (\varepsilon \vert \eta ) =& \frac{1}{\sqrt{2\pi} \sigma(Z)} \left\lbrace \Phi \left( \frac{ g\left( Z,\delta\right) \sigma_U \rho_U^\prime C^{-1}_\eta  D^{-1}_{\eta}\eta }{\lambda(Z) \sigma(Z)} -\frac{\lambda(Z) \left( \varepsilon -  \sigma_V \rho_V^\prime C^{-1}_\eta  D^{-1}_{\eta} \eta \right) }{\sigma(Z)}\right) \times \right. \\
& \quad \exp \left( -\frac{(\varepsilon - \sigma_V \rho_V^\prime C^{-1}_\eta  D^{-1}_{\eta}\eta + g\left( Z,\delta\right) \sigma_U \rho_U^\prime C^{-1}_\eta  D^{-1}_{\eta} \eta)^2}{2 \sigma^2(Z)}\right)\\
& \quad + \Phi \left( - \frac{ g\left( Z,\delta\right) \sigma_U \rho_U^\prime C^{-1}_\eta  D^{-1}_{\eta} \eta}{\lambda(Z) \sigma(Z)}  -\frac{\lambda(Z) \left( \varepsilon -  \sigma_V \rho_V^\prime C^{-1}_\eta  D^{-1}_{\eta} \eta \right) }{\sigma(Z)}\right) \times \\
& \quad \left. \exp \left( - \frac{(\varepsilon - \sigma_V \rho_V^\prime C^{-1}_\eta  D^{-1}_{\eta} \eta - g\left( Z,\delta\right) \sigma_U \rho_U^\prime C^{-1}_\eta  D^{-1}_{\eta}\eta)^2}{2 \sigma^2(Z)}\right) \right\rbrace,
\end{aligned}
\end{equation}
with
\begin{align*}
\lambda(Z) =& \frac{\tsigma_U(Z)}{\tsigma_V}, \quad \sigma^2(Z) = \tsigma^2_V + \tsigma^2_U(Z)\\
\tsigma^2_U(Z) =& (1- \rho_U^\prime C^{-1}_\eta \rho_U) \sigma^2_U (g \left( Z,\delta\right))^2,\\
\text{ and } & \tsigma^2_V =(1 -  \rho_V^\prime C^{-1}_\eta \rho_V)  \sigma^2_V
 \end{align*}

Upon the assumption of joint normality of the vector $\eta$, the joint density of $(\varepsilon,\eta)$ is therefore equal to
\begin{align}\label{eq:jointdens}
& f_{\varepsilon, \eta} (\varepsilon , \eta ) \nonumber \\
=& \frac{\vert D_\eta \vert^{-1}\vert C_\eta \vert^{-\frac{1}{2}}}{(2\pi)^{\frac{p + k + 1}{2}}  \sigma(Z)} \left\lbrace \Phi \left( \frac{ g\left( Z,\delta\right) \sigma_U \rho_U^\prime C^{-1}_\eta  D^{-1}_{\eta}\eta }{\lambda(Z) \sigma(Z)} -\frac{\lambda(Z) \left( \varepsilon -  \sigma_V \rho_V^\prime C^{-1}_\eta  D^{-1}_{\eta} \eta \right) }{\sigma(Z)}\right) \times \right. \nonumber \\
& \quad \exp \left( -\frac{(\varepsilon - \sigma_V \rho_V^\prime C^{-1}_\eta  D^{-1}_{\eta}\eta + g\left( Z,\delta\right) \sigma_U \rho_U^\prime C^{-1}_\eta  D^{-1}_{\eta} \eta)^2}{2 \sigma^2(Z)}  \right) \nonumber \\
& \quad + \Phi \left( - \frac{ g\left( Z,\delta\right) \sigma_U \rho_U^\prime C^{-1}_\eta  D^{-1}_{\eta} \eta}{\lambda(Z) \sigma(Z)}  -\frac{\lambda(Z) \left( \varepsilon -  \sigma_V \rho_V^\prime C^{-1}_\eta  D^{-1}_{\eta} \eta \right) }{\sigma(Z)}\right) \times \nonumber  \\
& \quad \left. \exp \left( - \frac{(\varepsilon - \sigma_V \rho_V^\prime C^{-1}_\eta  D^{-1}_{\eta} \eta - g\left( Z,\delta\right) \sigma_U \rho_U^\prime C^{-1}_\eta  D^{-1}_{\eta}\eta)^2}{2 \sigma^2(Z)} \right) \right\rbrace \exp \left( - \frac{1}{2} \eta^\prime  D^{-1}_\eta C^{-1}_\eta D^{-1}_\eta \eta\right) \nonumber \\
=& \frac{\vert D_\eta \vert^{-1}\vert C_\eta \vert^{-\frac{1}{2}}}{(2\pi)^{\frac{p + k + 1}{2}}  \sigma(Z)}  \exp \left( - \frac{1}{2} \eta^\prime  D^{-1}_\eta C^{-1}_\eta D^{-1}_\eta \eta\right)  \exp \left( -\frac{(\varepsilon - \sigma_V \rho_V^\prime C^{-1}_\eta  D^{-1}_{\eta}\eta + g\left( Z,\delta\right) \sigma_U \rho_U^\prime C^{-1}_\eta  D^{-1}_{\eta} \eta)^2}{2 \sigma^2(Z)}  \right)  \times \nonumber \\
& \quad \left\lbrace \Phi \left( \frac{ g\left( Z,\delta\right) \sigma_U\rho_U^\prime C^{-1}_\eta  D^{-1}_{\eta}\eta }{\lambda(Z) \sigma(Z)} -\frac{\lambda(Z) \left( \varepsilon -  \sigma_V \rho_V^\prime C^{-1}_\eta  D^{-1}_{\eta} \eta \right) }{\sigma(Z)}\right) \right. \nonumber \\
&\quad + \Phi \left( - \frac{ g\left( Z,\delta\right) \sigma_U\rho_U^\prime C^{-1}_\eta  D^{-1}_{\eta} \eta}{\lambda(Z) \sigma(Z)}  -\frac{\lambda(Z) \left( \varepsilon -  \sigma_V \rho_V^\prime C^{-1}_\eta  D^{-1}_{\eta} \eta \right) }{\sigma(Z)}\right) \times \nonumber \\
& \qquad \left.  \exp \left( \frac{2 (\varepsilon - \sigma_V \rho_V^\prime C^{-1}_\eta  D^{-1}_{\eta} \eta ) g\left( Z,\delta\right) \sigma_U \rho_U^\prime C^{-1}_\eta  D^{-1}_{\eta}\eta)}{\sigma^2(Z)} \right) \right\rbrace 
\end{align}

Replacing $\varepsilon = Y - m(X,\beta)$, and $\eta = R - W \gamma$, and taking logs, we obtain
\begin{align*}
\ell( \theta) =& -\frac{1}{2} \log \sigma^2(Z)  -  \log \vert D_\eta \vert  -\frac{1}{2}  \log \vert C_\eta \vert - \frac{1}{2} \left( R - W \gamma \right)^\prime D^{-1}_\eta C^{-1}_\eta D^{-1}_\eta \left( R - W \gamma \right) \\
& \quad - \frac{1}{2}  \frac{(Y - m(X,\beta)- \sigma_V \rho_V^\prime C^{-1}_\eta  D^{-1}_{\eta} \left( R - W \gamma \right) + g\left( Z,\delta\right) \sigma_U \rho_U^\prime C^{-1}_\eta  D^{-1}_{\eta}  \left( R - W \gamma \right))^2}{\sigma^2(Z)} \\
& \quad +\log \left\lbrace \Phi \left( \frac{g\left( Z,\delta\right) \sigma_U \rho_U^\prime C^{-1}_\eta  D^{-1}_{\eta}\left( R - W \gamma\right) }{\lambda(Z) \sigma(Z)} -\frac{\lambda(Z)}{\sigma(Z)} \left( Y - m(X,\beta) - \sigma_V \rho_V^\prime C^{-1}_\eta  D^{-1}_{\eta}\left( R - W \gamma\right)  \right)  \right)  \right. \\
& \quad + \Phi \left(  -\frac{g\left( Z,\delta\right) \sigma_U\rho_U^\prime C^{-1}_\eta  D^{-1}_{\eta}\left( R - W \gamma\right) }{\lambda(Z) \sigma(Z)} -\frac{\lambda(Z)}{\sigma(Z)} \left( Y - m(X,\beta) - \sigma_V \rho_V^\prime C^{-1}_\eta  D^{-1}_{\eta}\left( R - W \gamma\right) \right) \right) \times \\
& \quad \left. \exp \left( \frac{2 \left(Y - m(X,\beta) - \sigma_V \rho_V^\prime C^{-1}_\eta  D^{-1}_{\eta} \left( R - W \gamma \right) \right) g\left( Z,\delta\right) \sigma_U \rho_U^\prime C^{-1}_\eta  D^{-1}_{\eta}\left( R - W \gamma \right)}{\sigma^2(Z)} \right) \right\rbrace,
\end{align*}
\sloppy where we have omitted terms which do not depend on parameters, and $\theta = \left(\beta^\prime,\delta^\prime,\sigma^2_V,\sigma^2_U,\rho^\prime_V,\rho^\prime_U,\gamma^\prime,diag(D_\eta)^\prime,ve(C_\eta)^\prime \right)^\prime$. 

To simplify the notations below, let 
\begin{align*}
\omega =& Y - m(X,\beta) - \sigma_V \rho_V^\prime C^{-1}_\eta  D^{-1}_{\eta} \left( R - W \gamma \right)\\
\zeta =& g\left( Z,\delta\right) \sigma_U \rho_U^\prime C^{-1}_\eta  D^{-1}_{\eta}\left( R - W \gamma \right)\\
\xi =&  \frac{ g\left( Z,\delta\right) \sigma_U \rho_U^\prime C^{-1}_\eta  D^{-1}_{\eta}\left( R - W \gamma\right)  }{\lambda(Z) \sigma(Z)}  = \frac{\zeta}{\lambda(Z) \sigma(Z)}\\
\Psi =& \Phi \left( \xi - \frac{\lambda(Z)\omega}{\sigma(Z)} \right)  + \Phi \left(  -\xi -\frac{\lambda(Z)\omega}{\sigma(Z)}  \right) \exp \left( \frac{2 \omega \zeta}{\sigma^2(Z)} \right).
\end{align*}

The log-likelihood function can be finally written as
\begin{align*}
\ell(\theta) =& -\frac{1}{2} \log \sigma^2(Z) -  \log \vert D_\eta \vert -\frac{1}{2}  \log \vert C_\eta \vert \\
& \quad -  \frac{1}{2}\left( R - W \gamma \right)^\prime D^{-1}_\eta C^{-1}_\eta D^{-1}_\eta \left( R - W \gamma \right) - \frac{1}{2} \frac{(\omega + \zeta)^2}{\sigma^2(Z)} + \log \Psi.
\end{align*}
In the following, we use the facts that
\begin{equation}\label{eq:normdenseq}
\phi \left(  \xi -\frac{\lambda(Z)\omega}{\sigma(Z)}  \right)  = \phi \left(  -\xi -\frac{\lambda(Z)\omega}{\sigma(Z)}  \right) \exp \left( \frac{2\omega \zeta}{\sigma^2(Z)}\right),
\end{equation}
using the properties of the exponential function and the symmetry of the normal pdf, and
\begin{equation}\label{eq:normdenseq2}
\frac{\Phi \left(  -\xi -\frac{\lambda(Z)\omega}{\sigma(Z)}  \right) \exp \left( \frac{2 \omega \zeta}{\sigma^2(Z)} \right)}{\Psi} = 1 - \frac{\Phi \left(  \xi -\frac{\lambda(Z)\omega}{\sigma(Z)}  \right)}{\Psi}.
\end{equation}

We take the first order condition of the maximization problem with respect to $\rho_U$. Upon the conditions that the log-density is continuously differentiable and uniformly integrable on $\bar{\Theta}$, we can exchange expectation and differentiation. 

We thus have
\begin{equation} \label{eq:fscondrhou}
\begin{aligned}
\nabla_{\rho_U} \ell(\theta) =&  -\frac{1}{2} \frac{\nabla_{\rho_U} \sigma^2(Z)}{\sigma^2(Z)} - \left\lbrace \frac{\nabla_{\rho_U} \zeta (\zeta - \omega)}{\sigma^2(Z)} - \frac{\nabla_{\rho_U} \sigma^2(Z)}{2\sigma^4(Z)} (\zeta - \omega)^2 \right\rbrace \\
& \quad - \frac{2}{\Psi} \left\lbrace \left( \frac{\omega \nabla_{\rho_U} \lambda(Z)}{\sigma(Z)} - \frac{\nabla_{\rho_U} \sigma^2(Z) \lambda(Z) \omega}{2 \sigma^3(Z)} \right) \phi \left( \xi - \frac{\lambda(Z)\omega}{\sigma(Z)} \right) \right. \\
& \quad +\left.  \omega \Phi \left(  \xi -\frac{\lambda(Z)\omega}{\sigma(Z)}  \right) \left( \frac{\nabla_{\rho_U} \zeta}{\sigma^2 (Z)} - \frac{\zeta \nabla_{\rho_U} \sigma^2 (Z)}{\sigma^4 (Z)} \right)\right\rbrace = 0,
\end{aligned}
\end{equation}
where 
\begin{align*}
\nabla_{\rho_U} \sigma^2(Z)  =& -2C^{-1}_\eta \rho_U \sigma^2_U (g \left( Z,\delta\right))^2,\\
\nabla_{\rho_U} \lambda(Z)  =& -\frac{C^{-1}_\eta \rho_U \sigma^2_U (g \left( Z,\delta\right))^2}{\tsigma_V \tsigma^3_U(Z)},\\
\nabla_{\rho_U} \zeta =& C_\eta^{-1} D_\eta^{-1} \left( R - W \gamma \right) \sigma_U g(Z,\delta).
\end{align*}

To prove part (i) of the Proposition, it suffices to notice that, when $\rho_U = \Z$, then $\nabla_{\rho_U} \sigma^2(Z) = 0$, $\nabla_{\rho_U} \lambda(Z) = 0$, and $\xi = \zeta = 0$. Therefore 
\[
\Psi =  2 \Phi \left(-\frac{\lambda(Z)\omega}{\sigma(Z)}  \right),
\]
and 
\[
\left. \nabla_{\rho_U}\ell(\theta) \right\vert_{\rho_{U,0} = \Z}= \left\lbrace \frac{\omega \nabla_{\rho_U} \zeta }{\sigma^2(Z)} - \frac{\omega \nabla_{\rho_U} \zeta }{\sigma^2(Z)}  \right\rbrace,
\]
which is identically equal to zero. 

To prove part (ii), let $\rho_{U,0}$ be a solution to
\[
E \left[ \nabla_{\rho_U} \ell(\theta_0) \right] = \Z,
\]
with $\theta_{0} = (\theta^\prime_{1,0},\rho^\prime_{U,0})^\prime$. Further let $\xi_0$, $\zeta_0$, and $\omega_0$, be the values of $\xi$, $\zeta$ and $\omega$ at $\theta_0$, respectively. Let us evaluate the first order condition in \eqref{eq:fscondrhou} at $\rho_U = -\rho_{U,0}$. Notice that $\xi \vert_{-\rho_{U,0}} = -\xi_0$, $\zeta \vert_{-\rho_{U,0}} = -\zeta_0$, $\nabla_{\rho_U} \sigma^2(Z) \vert_{-\rho_{U,0}}= \nabla_{\rho_U} \sigma^2(Z) \vert_{\rho_{U,0}}$, and $\nabla_{\rho_U} \xi \vert_{-\rho_{U,0}}=\nabla_{\rho_U} \xi$. Therefore,
\begin{align*}
& E \left[ \nabla_{\rho_U}\ell(\theta_{1,0},-\rho_{U,0}) \right] =  \frac{1}{2} E \left[  \frac{\nabla_{\rho_U} \sigma_0^2(Z)}{\sigma_0^2(Z)}\right] + E \left[ \frac{\nabla_{\rho_U} \zeta_0 (\zeta_0 + \omega_0)}{\sigma_0^2(Z)} - \frac{\nabla_{\rho_U} \sigma_0^2(Z)}{2\sigma_0^4(Z)} (\zeta_0 + \omega_0)^2\right] \\
& \quad + E \left[ \frac{2 \left\lbrace \left(\frac{\omega_0 \nabla_{\rho_U} \lambda_0(Z)}{\sigma_0(Z)} - \frac{\nabla_{\rho_U} \sigma_0^2(Z) \lambda_0(Z) \omega_0}{2 \sigma_0^3(Z)} \right) \phi \left( -\xi_0 - \frac{\lambda_0(Z)\omega_0}{\sigma_0(Z)} \right) - \omega_0 \Phi \left(  -\xi_0 -\frac{\lambda_0(Z)\omega_0}{\sigma_0(Z)}  \right) \left( \frac{\nabla_{\rho_U} \zeta_0}{\sigma_0^2 (Z)} - \frac{\zeta_0 \nabla_{\rho_U} \sigma_0^2 (Z)}{\sigma_0^4 (Z)} \right)\right\rbrace}{\Phi \left(  -\xi_0 -\frac{\lambda_0(Z)\omega_0}{\sigma_0(Z)}  \right) + \Phi \left(  \xi_0 -\frac{\lambda_0(Z)\omega_0}{\sigma_0(Z)} \right) \exp \left( - \frac{2\omega_0 \zeta_0}{\sigma_0^2(Z)}\right)}\right]\\
&= \frac{1}{2} E \left[ \frac{\nabla_{\rho_U} \sigma_0^2(Z)}{\sigma_0^2(Z)} + \left\lbrace \frac{\nabla_{\rho_U} \zeta_0 (\zeta_0+ \omega_0)}{\sigma_0^2(Z)} - \frac{\nabla_{\rho_U} \sigma_0^2(Z)}{2\sigma_0^4(Z)} (\zeta_0 + \omega_0)^2 \right\rbrace \right]\\
& \quad + E \left[  \frac{2}{\Psi_0} \left\lbrace \left(\frac{\omega_0 \nabla_{\rho_U} \lambda_0(Z)}{\sigma_0(Z)} - \frac{\nabla_{\rho_U} \sigma_0^2(Z) \lambda_0(Z) \omega_0}{2 \sigma_0^3(Z)} \right) \phi \left( -\xi_0 - \frac{\lambda_0(Z)\omega_0}{\sigma_0(Z)} \right)\exp \left( - \frac{2\omega_0 \zeta_0}{\sigma_0^2(Z)}\right) \right.  \right.\\
& \quad \left. \left. - \omega_0 \Phi \left(  -\xi_0 -\frac{\lambda_0(Z)\omega_0}{\sigma_0(Z)}  \right) \exp \left( - \frac{2\omega_0 \zeta_0}{\sigma_0^2(Z)}\right) \left( \frac{\nabla_{\rho_U} \zeta_0}{\sigma_0^2 (Z)} - \frac{\zeta_0 \nabla_{\rho_U} \sigma_0^2 (Z)}{\sigma_0^4 (Z)} \right)\right\rbrace \right]\\
&= \frac{1}{2} E \left[ \frac{\nabla_{\rho_U} \sigma_0^2(Z)}{\sigma_0^2(Z)} + \frac{\nabla_{\rho_U} \zeta_0 (\zeta_0 - \omega_0)}{\sigma_0^2(Z)} - \frac{\nabla_{\rho_U} \sigma_0^2(Z)}{2\sigma_0^4(Z)} (\zeta_0 - \omega_0)^2 \right] \\
& \quad + E \left[ \frac{2}{\Psi} \left\lbrace \left(\frac{\omega_0 \nabla_{\rho_U} \lambda_0(Z)}{\sigma_0(Z)} - \frac{\nabla_{\rho_U} \sigma_0^2(Z) \lambda_0(Z) \omega_0}{2 \sigma_0^3(Z)} \right) \phi \left( \xi_0 - \frac{\lambda_0(Z)\omega_0}{\sigma_0(Z)} \right) \right. \right. \\
& \quad \left. \left. + \omega_0 \Phi \left(  \xi_0 -\frac{\lambda_0(Z)\omega_0}{\sigma_0(Z)}  \right) \left( \frac{\nabla_{\rho_U} \zeta_0}{\sigma_0^2 (Z)} - \frac{\zeta_0 \nabla_{\rho_U} \sigma_0^2 (Z)}{\sigma_0^4 (Z)} \right)\right\rbrace \right] = - E \left[ \nabla_{\rho_U} \ell( \theta_{0}) \right] = \Z,
\end{align*}
where the last step follows using the identities in equations \eqref{eq:normdenseq} and \eqref{eq:normdenseq2}. This proves part (ii). 

Finally, we show that the first derivative is not identically equal to $0$ when some (but not all) the components of $\rho_U$ are equal to $0$. Let $\rho_U = (\rho^\prime_{U1},\rho^\prime_{U2})^\prime$, where $\rho_{U1} \in \R^p$ and $\rho_{U2} \in \R^k$, for notational simplicity, and 
\[
C_\eta = \begin{bmatrix} C_{\eta,11} & C_{\eta,12} \\ C^\prime_{\eta,12}  & C_{\eta,22}\end{bmatrix}, \quad D_\eta = \begin{bmatrix}D_{\eta,1} & 0 \\ 0 &D_{\eta,2} \end{bmatrix}, \quad \eta =\begin{bmatrix} \eta_1 \\ \eta_2 \end{bmatrix}
\]
where $D_{\eta,1}$ and $D_{\eta,2}$ are diagonal matrices. Finally,
\begin{align*}
C^{-1}_\eta =& \begin{bmatrix} C^{-1}_{\eta,11} +  C^{-1}_{\eta,11} C_{\eta,12} B_{\eta,22} C^\prime_{\eta,12} & - C^{-1}_{\eta,11} C_{\eta,12} B_{\eta,22} \\ - B_{\eta,22} C^\prime_{\eta,12}  C^{-1}_{\eta,11}  & B_{\eta,22}\end{bmatrix} \\
=& \begin{bmatrix} C^{-1}_{\eta,11} & 0 \\ 0 & 0 \end{bmatrix} + \begin{bmatrix} - C^{-1}_{\eta,11}  C_{\eta,12}  \\ I_{k} \end{bmatrix} B_{\eta,22} \begin{bmatrix} -C^\prime_{\eta,12} C^{-1}_{\eta,11} & I_k \end{bmatrix}\\
=& \begin{bmatrix} C^{-1}_{\eta,11} & 0 \\ 0 & 0 \end{bmatrix} + P^\prime_\eta B_{\eta,22} P_\eta,
\end{align*}
with $B_{\eta,22} = \left( C_{\eta,22} - C^\prime_{\eta,12} C^{-1}_{\eta,11} C_{\eta,12}\right)^{-1}$, and $I_k$ the identity matrix of dimension $k$. 

Whenever $\rho_{U1,0} = \Z$, we have that 
\begin{equation} \label{eq:fscondrhou1}
\begin{aligned}
& \left. \nabla_{\rho_{U1}} \ell(\theta)\right\vert_{\rho_{U1} = \Z} =  - C^{-1}_{\eta,11} C_{\eta,12} B_{\eta,22}\frac{\rho_{U2} \sigma^2_U g^2(Z,\delta)}{\sigma^2(Z)} - C^{-1}_{\eta,11} D^{-1}_{\eta,1} \eta_1 \sigma_U g(Z,\delta) \frac{(\zeta - \omega)}{\sigma^2(Z)} \\
& \quad  + C^{-1}_{\eta,11} C_{\eta,12} B_{\eta,22} P_\eta  D^{-1}_{\eta} \eta \sigma_U g(Z,\delta) \frac{(\zeta - \omega)}{\sigma^2(Z)}+ C^{-1}_{\eta,11} C_{\eta,12} B_{\eta,22}\frac{\rho_{U2} \sigma^2_U g^2(Z,\delta)}{\sigma^4(Z)}(\zeta - \omega)^2 \\
& \quad  - \frac{2\omega}{\Psi} \left\lbrace \phi \left(\xi - \frac{\lambda(Z)}{\sigma(Z)}\omega\right)  \left[ C^{-1}_{\eta,11} C_{\eta,12} B_{\eta,22}\frac{\rho_{U2} \sigma^2_U g^2(Z,\delta)}{\sigma(Z) \tsigma_V \tsigma_U^3(Z)}  -  C^{-1}_{\eta,11} C_{\eta,12} B_{\eta,22}\frac{\rho_{U2} \sigma^2_U g^2(Z,\delta) \lambda(Z)}{\sigma^3(Z) } \right] \right.\\
& \quad + \Phi \left(\xi - \frac{\lambda(Z)}{\sigma(Z)}\omega\right)\left[ C^{-1}_{\eta,11} D^{-1}_{\eta,1} \eta_1 \frac{\sigma_U g(Z,\delta)}{\sigma^2(Z)} - C^{-1}_{\eta,11} C_{\eta,12} B_{\eta,22}P_\eta  D^{-1}_{\eta} \eta \frac{\sigma_U g(Z,\delta)}{\sigma^2(Z)} \right. \\
& \quad \left. \left. +  C^{-1}_{\eta,11} C_{\eta,12} B_{\eta,22}\rho_{U2}  \frac{2\zeta \sigma^2_U g^2(Z,\delta)}{\sigma^4(Z)} \right] \right\rbrace,
\end{aligned}
\end{equation}
and 
\begin{equation} \label{eq:fscondrhou2}
\begin{aligned}
\left. \nabla_{\rho_{U2}} \ell(\theta)\right\vert_{\rho_{U1} = \Z} =&  B_{\eta,22}\frac{\rho_{U2} \sigma^2_U g^2(Z,\delta)}{\sigma^2(Z)} - B_{\eta,22} P_\eta D^{-1}_{\eta} \eta \sigma_U g(Z,\delta) \frac{(\zeta - \omega)}{\sigma^2(Z)}- B_{\eta,22}\frac{\rho_{U2} \sigma^2_U g^2(Z,\delta)}{\sigma^4(Z)}(\zeta - \omega)^2 \\
& \quad  - \frac{2\omega}{\Psi} \left\lbrace \phi \left(\xi - \frac{\lambda(Z)}{\sigma(Z)}\omega\right)  \left[ - B_{\eta,22}\frac{\rho_{U2} \sigma^2_U g^2(Z,\delta)}{\sigma(Z) \tsigma_V \tsigma_U^3(Z)}  + B_{\eta,22}\frac{\rho_{U2} \sigma^2_U g^2(Z,\delta) \lambda(Z)}{\sigma^3(Z) } \right] \right.\\
& \quad + \Phi \left(\xi - \frac{\lambda(Z)}{\sigma(Z)}\omega\right)\left[ B_{\eta,22} P_\eta D^{-1}_{\eta} \eta \frac{\sigma_U g(Z,\delta)}{\sigma^2(Z)} +  B_ {\eta,22}\rho_{U2}  \frac{2\zeta \sigma^2_U g^2(Z,\delta)}{\sigma^4(Z)} \right\rbrace. 
\end{aligned}
\end{equation}

Notice that 
\begin{align*}
& \left. \nabla_{\rho_{U1}} \ell(\theta)\right\vert_{\rho_{U1} = \Z} =  - C^{-1}_{\eta,11} C_{\eta,12} \left. \nabla_{\rho_{U2}} \ell(\theta)\right\vert_{\rho_{U1} = \Z}  - C^{-1}_{\eta,11} D^{-1}_{\eta,1} \eta_1 \sigma_U g(Z,\delta) \frac{(\zeta - \omega)}{\sigma^2(Z)} \\
& \quad  - \frac{2\omega}{\Psi} \Phi \left(\xi - \frac{\lambda(Z)}{\sigma(Z)}\omega\right) C^{-1}_{\eta,11} D^{-1}_{\eta,1} \eta_1 \frac{\sigma_U g(Z,\delta)}{\sigma^2(Z)}, 
\end{align*}

which by evaluating the expression at $\theta_0$, and taking expectations gives
\begin{align*}
& E\left[ \nabla_{\rho_{U1}} \ell(\theta_0) \right] =  - C^{-1}_{\eta,11} C_{\eta,12} E\left[ \nabla_{\rho_{U2}} \ell(\theta_0) \right]  - C^{-1}_{\eta,11} D^{-1}_{\eta,1} E\left[ \eta_1 \sigma_{U} g(Z,\delta) \frac{(\zeta - \omega)}{\sigma^2(Z)} \right] \\
& \quad  - 2 C^{-1}_{\eta,11} D^{-1}_{\eta,1} E\left[ \frac{\omega}{\Psi} \Phi \left(\xi - \frac{\lambda(Z)}{\sigma(Z)}\omega\right) \eta_1 \frac{\sigma_U g(Z,\delta)}{\sigma^2(Z)}\right].
\end{align*}
where $E\left[ \nabla_{\rho_{U2}} \ell(\theta_0) \right] = \Z$, by definition. 

We now show that the remaining two terms are equal to $0$ in expectation. The conditional pdf of $\varepsilon$ given $\eta$ can be written as 
\[
f_{\varepsilon \vert \eta} (\varepsilon \vert \eta ) = \frac{1}{\sigma(Z)} \phi \left( \frac{\omega + \zeta}{\sigma(Z)}\right) \Psi. 
\]
As this is a mixture of two conditional extended skew normal distributions, we have that 
\begin{align*}
E \left[ \omega \vert \eta \right] =& \Phi \left( \frac{\tsigma^2_V \zeta}{\tsigma_U(Z) \sigma^2(Z)}  \right)\left( - \zeta +  \tsigma_U (Z) \frac{\phi\left( \frac{\tsigma^2_V \zeta}{\tsigma_U(Z) \sigma^2(Z)}  \right) }{\Phi\left( \frac{\tsigma^2_V \zeta}{\tsigma_U(Z) \sigma^2(Z)}   \right)}\right) + \left( 1  - \Phi \left( \frac{\tsigma^2_V \zeta}{\tsigma_U(Z) \sigma^2(Z)}   \right) \right)\left(\zeta +  \tsigma_U (Z) \frac{\phi\left( - \frac{\tsigma^2_V \zeta}{\tsigma_U(Z) \sigma^2(Z)}  \right) }{1 - \Phi\left( \frac{\zeta}{\tsigma_U(Z)}  \right)}\right)\\
=& 2  \tsigma_U (Z) \phi\left( \frac{\tsigma^2_V \zeta}{\tsigma_U(Z) \sigma^2(Z)}   \right)+  \left( 1 - 2 \Phi\left( \frac{\tsigma^2_V \zeta}{\tsigma_U(Z) \sigma^2(Z)}  \right) \right) \zeta,
\end{align*}
and
\begin{align*}
E \left[ \frac{\omega}{\Psi} \Phi \left(\xi - \frac{\lambda(Z)}{\sigma(Z)}\omega\right) \vert \eta \right] =& - \Phi\left( \frac{\tsigma^2_V \zeta}{\tsigma_U(Z) \sigma^2(Z)}  \right) \zeta + \tsigma_U(Z) \phi\left( \frac{\tsigma^2_V \zeta}{\tsigma_U(Z) \sigma^2(Z)}   \right),
\end{align*}
where both results follow from the Moment Generating Function of the extended skew-normal distribution given in \citet[][p. 35-36]{azzalini2013}.

Finally, 
\begin{align*}
- E & \left[ \eta_1 (\zeta - \omega)\frac{\sigma_{U} g(Z,\delta)}{\sigma^2(Z)} \right] = 2 E\left[ \eta_1 \left( \tsigma_U (Z) \phi\left( \frac{\tsigma^2_V \zeta}{\tsigma_U(Z) \sigma^2(Z)} \right) - \Phi\left( \frac{\tsigma^2_V \zeta}{\tsigma_U(Z) \sigma^2(Z)}  \right) \zeta \right)\frac{\sigma_{U} g(Z,\delta)}{\sigma^2(Z)} \right],
\end{align*}
and
\begin{align*}
- 2E & \left[ \frac{\omega}{\Psi} \Phi \left(\xi - \frac{\lambda(Z)}{\sigma(Z)}\omega\right) \eta_1 \frac{\sigma_U g(Z,\delta)}{\sigma^2(Z)}\right] = - 2 E\left[ \eta_1 \left( \tsigma_U(Z) \phi\left(\frac{\tsigma^2_V \zeta}{\tsigma_U(Z) \sigma^2(Z)} \right) - \Phi\left( \frac{\tsigma^2_V \zeta}{\tsigma_U(Z) \sigma^2(Z)}  \right) \zeta \right)  \frac{\sigma_{U} g(Z,\delta)}{\sigma^2(Z)} \right],
\end{align*}
and their sum is therefore equal to $0$. This concludes the proof.

\subsection{Proof of Proposition \ref{prop:identif2}}
%%% Second derivative wrt rhou
Notice that $\nabla^2_{\rho_U\rho_U^\prime} \zeta = \Z$. The second derivative of the log-likelihood function wrt $\rho_U$ is thus equal to
\begin{align*}
\nabla^2_{\rho_U \rho^\prime_U}  \ell(\theta) =& -\frac{1}{2} \left\lbrace \frac{\nabla^2_{\rho_U \rho^\prime_U} \sigma^2 (Z)}{\sigma^2(Z)} - \frac{\nabla_{\rho_U} \sigma^2 ( Z) \nabla_{\rho^\prime_U} \sigma^2 ( Z)}{\sigma^4(Z)} \right\rbrace \\
& \quad -\left\lbrace \frac{\nabla_{\rho_U} \zeta \nabla_{\rho^\prime_U}\zeta }{\sigma^2(Z)}  - \frac{\nabla_{\rho_U} \zeta (\zeta - \omega) \nabla_{\rho^\prime_U}\sigma^2(Z) }{\sigma^4(Z)} \right\rbrace \\
& \quad + \left\lbrace  \frac{\nabla^2_{\rho_U \rho^\prime_U} \sigma^2(Z)  (\zeta - \omega)^2 }{2 \sigma^4(Z)}  +  \frac{\nabla_{\rho_U} \sigma^2(Z)(\zeta - \omega)  \nabla_{\rho^\prime_U}\zeta }{\sigma^4(Z)}  - \frac{\nabla_{\rho_U} \sigma^2(Z) (\zeta - \omega)^2 \nabla_{\rho^\prime_U}\sigma^2(Z) }{\sigma^6(Z)} \right\rbrace \\
& \quad -\frac{2\omega }{\Psi} \left\lbrace \left[ \frac{\nabla^2_{\rho_U \rho^\prime_U} \lambda(Z)}{\sigma(Z)} - \frac{\nabla_{\rho_U} \lambda(Z) \nabla_{\rho^\prime_U} \sigma^2(Z)}{2\sigma^3(Z)} - \frac{ \nabla^2_{\rho_U \rho^\prime_U} \sigma^2(Z) \lambda(Z)}{2\sigma^3(Z)} \right. \right. \\
& \quad - \frac{\nabla_{\rho_U} \sigma^2(Z)\nabla_{\rho^\prime_U}  \lambda(Z)}{2\sigma^3(Z)} + \frac{3\lambda(Z) \nabla_{\rho_U} \sigma^2 ( Z) \nabla_{\rho^\prime_U} \sigma^2 ( Z)}{4 \sigma^5(Z)} \\
& \quad +\left( \frac{\nabla_{\rho_U} \zeta}{\sigma^2 (Z)} - \frac{\zeta \nabla_{\rho_U} \sigma^2 (Z)}{\sigma^4 (Z)} \right) \left( \nabla_{\rho^\prime_U} \xi - \frac{\nabla_{\rho^\prime_U}  \lambda (Z) \omega}{\sigma (Z)} + \frac{\lambda(Z) \nabla_{\rho^\prime_U}  \sigma^2 (Z) \omega}{2 \sigma^3 (Z)} \right) \\
& \quad - \left( \xi - \frac{\lambda(Z)\omega}{\sigma(Z)}\right) \left( \frac{\nabla_{\rho_U} \lambda(Z)}{\sigma(Z)} - \frac{\nabla_{\rho_U} \sigma^2(Z) \lambda(Z)}{2 \sigma^3(Z)} \right) \times \\
& \qquad \left. \left( \nabla_{\rho^\prime_U} \xi - \frac{\nabla_{\rho^\prime_U}  \lambda (Z) \omega}{\sigma (Z)} + \frac{\lambda(Z) \nabla_{\rho^\prime_U}  \sigma^2 (Z) \omega}{2 \sigma^3 (Z)} \right) \right] \phi \left( \xi - \frac{\lambda(Z) \omega}{\sigma(Z)}\right)  \\
& \quad - \left.  \left(  \frac{2\nabla_{\rho_U} \zeta \nabla_{\rho^\prime_U} \sigma^2(Z)}{\sigma^4(Z)} + \frac{\zeta \nabla^2_{\rho_U \rho^\prime_U} \sigma^2(Z)}{\sigma^4(Z)}  - \frac{2 \zeta \nabla_{\rho_U} \sigma^2(Z)\nabla_{\rho^\prime_U} \sigma^2(Z)}{\sigma^6(Z)}  \right) \Phi \left( \xi - \frac{\lambda(Z) \omega}{\sigma(Z)}\right)  \right\rbrace \\
& \quad + \frac{4\omega^2}{\Psi^2} \left\lbrace  \left( \frac{\nabla_{\rho_U} \lambda(Z)}{\sigma(Z)} - \frac{\nabla_{\rho_U} \sigma^2(Z) \lambda(Z)}{2 \sigma^3(Z)} \right) \phi \left( \xi - \frac{\lambda(Z)\omega}{\sigma(Z)} \right) \right. \\
& \quad +\left.  \Phi \left(  \xi -\frac{\lambda(Z)\omega}{\sigma(Z)}  \right) \left( \frac{\nabla_{\rho_U} \zeta}{\sigma^2 (Z)} - \frac{\zeta \nabla_{\rho_U} \sigma^2 (Z)}{\sigma^4 (Z)} \right) \right\rbrace \times \\
& \qquad\left\lbrace \left( -\frac{\nabla_{\rho^\prime_U} \lambda(Z)}{\sigma(Z)} + \frac{\nabla_{\rho^\prime_U} \sigma^2(Z) \lambda(Z)}{2 \sigma^3(Z)} \right) \phi \left( \xi - \frac{\lambda(Z)\omega}{\sigma(Z)} \right) \right. \\
&\qquad \left. + \Phi \left( -\xi - \frac{\lambda(Z) \omega}{\sigma(Z)}\right) \exp \left( \frac{2\omega \zeta}{\sigma^2(Z)}\right) \left( \frac{\nabla_{\rho^\prime_U} \zeta}{\sigma^2 (Z)} - \frac{\zeta \nabla_{\rho^\prime_U} \sigma^2 (Z)}{\sigma^4 (Z)} \right)  \right\rbrace,
\end{align*}
where 
\begin{align*}
\nabla^2_{\rho_U \rho^\prime_U} \sigma^2(Z)  =& -2C^{-1}_\eta \sigma^2_U (g \left( Z,\delta\right))^2,\\
\nabla^2_{\rho_U \rho^\prime_U} \lambda(Z)  =& -\frac{1}{\tsigma_V}\left[ \frac{C^{-1}_\eta \sigma^2_U (g \left( Z,\delta\right))^2}{\tsigma^3_U(Z)} - \frac{3 C^{-1}_\eta \rho_U \sigma^4_U (g \left( Z,\delta\right))^4 \rho^\prime_U C^{-1}_\eta}{2\tsigma^5_U(Z)}  \right].
\end{align*}

Whenever $\rho_U = \Z$,
\begin{align*}
\left. \nabla^2_{\rho_U \rho^\prime_U} \ell(\theta) \right\vert_{\rho_{U} = \Z}=& \frac{C^{-1}_\eta \sigma_U^2 g^2(Z,\delta)}{\sigma^2(Z)} - \frac{\nabla_{\rho_U} \zeta \nabla_{\rho^\prime_U} \zeta}{\sigma^2(Z)} - \frac{C^{-1}_\eta \sigma_U^2 g^2(Z,\delta) \omega^2}{\sigma^4(Z)} \\
& \quad - \frac{\phi\left( -\frac{\lambda(Z)}{\sigma(Z)}\omega\right) \omega}{\Phi\left( -\frac{\lambda(Z)}{\sigma(Z)}\omega\right) } \left\lbrace -  \frac{C^{-1}_\eta \sigma_U^2 g^2(Z,\delta)\lambda(Z)}{\sigma(Z)\tsigma^4_U(Z)} + \frac{C^{-1}_\eta \sigma_U^2 g^2(Z,\delta)\lambda(Z)}{\sigma^3(Z)}  \right. \\
& \quad \left.+ \frac{\nabla_{\rho_U} \zeta \nabla_{\rho^\prime_U} \zeta}{\lambda(Z) \sigma^3(Z)} \right\rbrace + \frac{\nabla_{\rho_U} \zeta \nabla_{\rho^\prime_U} \zeta}{\sigma^4(Z)} \omega^2.
\end{align*}

Notice that, when $\rho_U= \Z$,
\begin{align*}
f_{\varepsilon \vert \eta } ( \varepsilon \vert \eta) = \frac{2}{\sigma(Z)} \Phi \left( -\frac{\lambda(Z)}{\sigma(Z)}\omega\right) \phi\left( \frac{\omega}{\sigma(Z)}\right),
\end{align*}
which implies that 
\begin{align*}
E\left[ \varepsilon \vert \eta \right] =& \sigma_V \rho^\prime_V C^{-1}_\eta D^{-1}_\eta \eta - \tsigma_U(Z) \sqrt{\frac{2}{\pi}} \\
Var \left[ \varepsilon \vert \eta \right] =&\sigma^2(Z) - \tsigma^2_U(Z) \frac{2}{\pi}. 
\end{align*}
Therefore,
\begin{align*}
E\left[ \omega^2 \vert \eta \right] =& Var \left[ \varepsilon \vert \eta \right] +  \left( E\left[ \varepsilon \vert \eta \right]  \right)^2 - 2E\left[ \varepsilon \vert \eta \right]   \sigma_V \rho^\prime_V C^{-1}_\eta D^{-1}_\eta \eta  + \left(  \sigma_V \rho^\prime_V C^{-1}_\eta D^{-1}_\eta \eta \right)^2 \\
=& \sigma^2(Z) - \tsigma^2_U(Z) \frac{2}{\pi} +\left(  \sigma_V \rho^\prime_V C^{-1}_\eta D^{-1}_\eta \eta \right)^2 + \tsigma^2_U(Z) \frac{2}{\pi}  - 2 \left(  \sigma_V \rho^\prime_V C^{-1}_\eta D^{-1}_\eta \eta \right) \tsigma_U(Z) \sqrt{\frac{2}{\pi}}  \\
& \quad - 2 \left(  \sigma_V \rho^\prime_V C^{-1}_\eta D^{-1}_\eta \eta \right)^2 + 2 \left(  \sigma_V \rho^\prime_V C^{-1}_\eta D^{-1}_\eta \eta \right) \tsigma_U(Z) \sqrt{\frac{2}{\pi}} +\left(  \sigma_V \rho^\prime_V C^{-1}_\eta D^{-1}_\eta \eta \right)^2\\
=& \sigma^2(Z),
\end{align*}
and 
\begin{align*}
E\left[ \frac{\phi\left( -\frac{\lambda(Z)}{\sigma(Z)}\omega\right) \omega}{\Phi\left( -\frac{\lambda(Z)}{\sigma(Z)}\omega\right) } \vert \eta \right] =& \frac{2}{\sigma(Z)}  \int  \omega \phi\left( -\frac{\lambda(Z)}{\sigma(Z)}\omega\right) \phi\left( \frac{\omega}{\sigma(Z)}\right) d\varepsilon \\
=& \frac{2\tsigma_V}{\sqrt{2\pi}\sigma(Z)} \int \frac{\left( \varepsilon -  \sigma_V \rho^\prime_V C^{-1}_\eta D^{-1}_\eta \eta \right) }{\tsigma_V} \phi\left( \frac{\varepsilon -  \sigma_V \rho^\prime_V C^{-1}_\eta D^{-1}_\eta \eta}{\tsigma_V}\right)d \varepsilon = 0,
\end{align*}
as the integral can be seen as the mean of a centered normally distributed random variable. 

Therefore, 
\[
E \left[ \nabla^2_{\rho_U \rho^\prime_U} \ell(\theta_0) \right] = \Z, 
\]
where the result follows from the law of iterated expectations, and the conditional independence of $\varepsilon$ and $R$ given $\eta$ from Assumption \ref{ass:cindepeta}.

\subsection{Proof of Theorem \ref{thm:asnorm}}

Our proof is based on \citet{andrews1999} and \citet{rot2000}. In particular, we show that the log-likelihood admits a quadratic expansion at $\theta_0 = (\theta^\prime_{1,0},0^\prime)^\prime$, and then directly claim Theorems 2-3 and 4 of \citet{andrews1999}. The likelihood function is infinitely differentiable wrt $\theta$ at $\theta_0$. We further assume the following

\begin{assumption} \label{assapp:diff}
The derivatives of the log-likelihood function up to the sixth order are square-integrable wrt the distribution of the data.
\end{assumption}

The reason to require derivatives up to the sixth order to exist is that the log-likelihood function is an even function of $\rho_U$, as shown in Proposition \ref{prop:identif1}. Therefore in the MacLaurin series of $\ell_n(\theta_0)$ all odd derivatives wrt $\rho_U$ are equal to $0$. Given Assumption \ref{assapp:diff}, and the iid assumption, we can then use the weak law of large numbers (WLLN) and the central limit theorem (CLT) on sample objects.  

Let $\hat{\theta}_n$ be the maximum likelihood estimator of $\theta^\ast_0$. Then 
\begin{align*}
\ell_n(\hat{\theta}_n) =& \ell_n(\theta^\ast_0) + \nabla_{\theta_1} \ell_n(\theta^\ast_0)^\prime \left(\hat\theta_{1,n} - \theta_{1,0} \right) +  \nabla_{\rho_U} \ell_n(\theta^\ast_0)^\prime \hat{\rho}_{U,n} \\
&\quad + \frac{1}{2}\left(\hat\theta_{1,n} - \theta_{1,0} \right)^\prime  \nabla^2_{\theta_1 \theta_1^\prime}  \ell_n(\theta^\ast_0) \left(\hat\theta_{1,n} - \theta_{1,0} \right) + \frac{1}{2}\hat\rho_{U,n}^\prime  \nabla^2_{\rho_U \rho_U^\prime}  \ell_n(\theta^\ast_0) \hat\rho_{U,n} \\
& \quad + 2 \hat\rho_{U,n}^\prime  \nabla^2_{\rho_U \theta_1^\prime}  \ell_n(\theta^\ast_0) \left(\hat\theta_{1,n} - \theta_{1,0} \right) + \frac{1}{3!} vec(\hat\rho_{U,n}\hat\rho_{U,n}^\prime) \nabla_{\rho^\prime_U} vec \left( \nabla^2_{\rho_U \rho_U^\prime}  \ell_n(\theta^\ast_0) \right)\hat\rho_{U,n} \\
& \quad + \frac{2}{3!} vec(\hat\rho_{U,n}\hat\rho_{U,n}^\prime)^\prime \nabla_{\theta^\prime_1} vec \left( \nabla^2_{\rho_U \rho_U^\prime}  \ell_n(\theta^\ast_0) \right)\left(\hat\theta_{1,n} - \theta_{1,0} \right) \\
& \quad + \frac{1}{4!}vec(\hat\rho_{U,n}\hat\rho_{U,n}^\prime)^\prime \left[\nabla^2_{vec(\rho_U \rho^\prime_U)} vec \left( \nabla^2_{\rho_U \rho_U^\prime}  \ell_n(\theta^\ast_0) \right) \right] vec(\hat\rho_{U,n}\hat\rho_{U,n}^\prime) \\
& \quad + \Vert \hat\theta_{1,n} - \theta_{1,0} \Vert^3 h_{\theta_1,n}(Y_i,X_i,Z_i) + \Vert \hat\rho_{U,n}\Vert^6 h_{\rho_U,n}(Y_i,X_i,Z_i),
\end{align*}
with $E\left[ h^2_{\theta_1,n}(Y_i,X_i,Z_i) \right] < \infty$ and $E\left[ h^2_{\rho_{U},n}(Y_i,X_i,Z_i) \right] < \infty$ by Assumption \ref{assapp:diff}. Therefore, by the consistency of our maximum likelihood estimator, the remainder is $o_P(1)$. By the proof of Proposition \ref{prop:identif1}, we know that the fist derivative with respect to $\rho_U$ at zero is identically equal to zero. This further implies that the first cross partial derivatives are also equal to zero, by the second Bartlett's identity. We also show in a Supplementary Appendix, that the third derivative wrt $\rho_U$ at zero is identically zero. Using these identities, 
\begin{align*}
\ell_n(\hat{\theta}_n) =& \ell_n(\theta^\ast_0) + \nabla_{\theta_1} \ell_n(\theta^\ast_0)^\prime \left(\hat\theta_{1,n} - \theta_{1,0} \right) \\
&\quad + \frac{1}{2}\left(\hat\theta_{1,n} - \theta_{1,0} \right)^\prime  \nabla^2_{\theta_1 \theta_1^\prime}  \ell_n(\theta^\ast_0) \left(\hat\theta_{1,n} - \theta_{1,0} \right) + \frac{1}{2}\hat\rho_{U,n}^\prime  \nabla^2_{\rho_U \rho_U^\prime}  \ell_n(\theta^\ast_0) \hat\rho_{U,n} \\
& \quad + \frac{2}{3!} vec(\hat\rho_{U,n}\hat\rho_{U,n}^\prime)^\prime \nabla_{\theta^\prime_1} vec \left( \nabla^2_{\rho_U \rho_U^\prime}  \ell_n(\theta^\ast_0) \right)\left(\hat\theta_{1,n} - \theta_{1,0} \right) \\
& \quad + \frac{1}{4!}vec(\hat\rho_{U,n}\hat\rho_{U,n}^\prime)^\prime \left[\nabla^2_{vec(\rho_U \rho^\prime_U)} vec \left( \nabla^2_{\rho_U \rho_U^\prime}  \ell_n(\theta^\ast_0) \right) \right] vec(\hat\rho_{U,n}\hat\rho_{U,n}^\prime) + o_P(1)\\
=& \ell_n(\theta^\ast_0) + \nabla_{\theta_1} \ell_n(\theta^\ast_0)^\prime \left(\hat\theta_{1,n} - \theta_{1,0} \right) \\
&\quad + \frac{1}{2}\left(\hat\theta_{1,n} - \theta_{1,0} \right)^\prime  \nabla^2_{\theta_1 \theta_1^\prime}  \ell_n(\theta^\ast_0) \left(\hat\theta_{1,n} - \theta_{1,0} \right) + \frac{1}{2}vec \left( \nabla^2_{\rho_U \rho_U^\prime} \ell_n(\theta^\ast_0)\right)^\prime vec(\hat\rho_{U,n}\hat\rho_{U,n}^\prime) \\
& \quad + \frac{2}{3!} vec(\hat\rho_{U,n}\hat\rho_{U,n}^\prime)^\prime \nabla_{\theta^\prime_1} vec \left( \nabla^2_{\rho_U \rho_U^\prime}  \ell_n(\theta^\ast_0) \right)\left(\hat\theta_{1,n} - \theta_{1,0} \right) \\
& \quad + \frac{1}{4!}vec(\hat\rho_{U,n}\hat\rho_{U,n}^\prime)^\prime \left[\nabla^2_{vec(\rho_U \rho^\prime_U)} vec \left( \nabla^2_{\rho_U \rho_U^\prime}  \ell_n(\theta^\ast_0) \right) \right] vec(\hat\rho_{U,n}\hat\rho_{U,n}^\prime) + o_P(1)\\
=& \ell_n(\theta^\ast_0)  + \left( \frac{\nabla_{\theta_1} \ell_n(\theta^\ast_0)}{\sqrt{n}} \right)^\prime \sqrt{n} \left(\hat\theta_{1,n} - \theta_{1,0} \right) \\
& \quad + \frac{1}{2}\sqrt{n}\left(\hat\theta_{1,n} - \theta_{1,0} \right)^\prime \frac{\nabla^2_{\theta_1 \theta_1^\prime}  \ell_n(\theta^\ast_0)}{n}\sqrt{n}\left(\hat\theta_{1,n} - \theta_{1,0} \right) + \frac{1}{2}\frac{vec \left( \nabla^2_{\rho_U \rho_U^\prime} \ell_n(\theta^\ast_0)\right)^\prime}{\sqrt{n}} \sqrt{n}vec(\hat\rho_{U,n}\hat\rho_{U,n}^\prime) \\
& \quad + \frac{2}{3!} \sqrt{n} vec(\hat\rho_{U,n}\hat\rho_{U,n}^\prime)^\prime \frac{\nabla_{\theta^\prime_1} vec \left( \nabla^2_{\rho_U \rho_U^\prime}  \ell_n(\theta^\ast_0) \right)}{n}\sqrt{n}\left(\hat\theta_{1,n} - \theta_{1,0} \right) \\
& \quad + \frac{1}{4!}\sqrt{n} vec(\hat\rho_{U,n}\hat\rho_{U,n}^\prime)^\prime \left[\frac{\nabla^2_{vec(\rho_U \rho^\prime_U)} vec \left( \nabla^2_{\rho_U \rho_U^\prime}  \ell_n(\theta^\ast_0) \right) }{n}\right] \sqrt{n}vec(\hat\rho_{U,n}\hat\rho_{U,n}^\prime) + o_P(1).
\end{align*}

Let
\[
Z = \begin{pmatrix} Z_{\theta_1}\\  Z_{\rho_U \rho_U}\end{pmatrix} \sim N\left(0, \mathcal{I}_1^{-} \right). 
\]
Recall that, by the properties of the Moore-Penrose pseudo inverse, $\mathcal{I}_1^{-} \mathcal{I}_1 \mathcal{I}_1^{-} = \mathcal{I}_1^{-}$, and $\mathcal{I}_1 \mathcal{I}_1^{-} \mathcal{I}_1 = \mathcal{I}_1$.

By Corollary 1 in \citet{rot2000}, the WLLN and the CLT, we have that 
\begin{align*}
-& \frac{\nabla^2_{\theta_1 \theta_1^\prime}  \ell_n(\theta^\ast_0)}{n} \xrightarrow{p} \mathcal{I}_{\theta_1}\\
-& \frac{\nabla^2_{vec(\rho_U \rho^\prime_U)} vec \left( \nabla^2_{\rho_U \rho_U^\prime}  \ell_n(\theta^\ast_0) \right) }{n} \xrightarrow{p} \frac{4!\mathcal{I}_{\rho_U \rho_U^\prime}}{2}\\
-& \frac{\nabla_{\theta^\prime_1} vec \left( \nabla^2_{\rho_U \rho_U^\prime}  \ell_n(\theta^\ast_0) \right)}{n} \xrightarrow{p} \frac{3!\mathcal{I}_{\theta_1 \rho_U\rho_U^\prime}}{2}\\
& \begin{pmatrix} \frac{\nabla_{\theta_1} \ell_n(\theta^\ast_0)}{\sqrt{n}} \\ \frac{vec \left( \nabla^2_{\rho_U \rho_U^\prime} \ell_n(\theta^\ast_0)\right)}{2\sqrt{n}} \end{pmatrix} \xrightarrow{d} \mathcal{I}_1 \begin{pmatrix} Z_{\theta_1} \\ Z_{\rho_U \rho_U}\end{pmatrix}.
\end{align*}

All the regularity conditions in \citet[][Assumptions 2-6]{andrews1999} are satisfied, so that $\sqrt{n}\left(\hat\theta_{1,n} - \theta_{1,0} \right) = \hat{\tau}_{\theta_1,n} + o_P(1)$, and $\sqrt{n}vec(\hat\rho_{U,n}\hat\rho_{U,n}^\prime) = \hat{\tau}_{\rho_U\rho_U^\prime,n}  + o_P(1)$, where $\hat{\tau}_{\theta_1,n} \xrightarrow{d} \hat{\tau}_{\theta_1}$ and $\hat{\tau}_{\rho_U\rho_U^\prime,n} \xrightarrow{d} \hat{\tau}_{\rho_U\rho_U^\prime}$ as $n \rightarrow \infty$. 
Thus
\begin{align*}
\ell_n(\hat{\theta}_n) =& \ell_n(\theta^\ast_0)  + Z^\prime_{\theta_1} \mathcal{I}_{\theta_1}  \hat{\tau}_{\theta_1,n} + Z^\prime_{\rho_U\rho_U^\prime} \mathcal{I}_{\rho_U\rho_U^\prime\theta_1}  \hat{\tau}_{\theta_1,n}-\frac{1}{2}\hat{\tau}^\prime_{\theta_1,n} \mathcal{I}_{\theta_1} \hat{\tau}_{\theta_1,n} \\
& \quad + Z^\prime_{\rho_U\rho_U^\prime} \mathcal{I}_{\rho_U\rho_U^\prime} \hat{\tau}_{\rho_U\rho_U^\prime,n} + Z^\prime_{\theta_1} \mathcal{I}_{\theta_1\rho_U\rho_U^\prime}\hat{\tau}_{\rho_U\rho_U^\prime,n} \\
& \quad - \hat{\tau}^\prime_{\rho_U\rho_U^\prime,n} \mathcal{I}_{\rho_U\rho_U^\prime\theta_1}  \hat{\tau}_{\theta_1,n} - \frac{1}{2} \hat{\tau}^\prime_{\rho_U\rho_U^\prime,n} \mathcal{I}_{\rho_U\rho_U^\prime\theta_1} \hat{\tau}_{\rho_U\rho_U^\prime,n} + o_P(1)\\
=& \ell_n(\theta^\ast_0)  + \frac{1}{2} Z^\prime \mathcal{I} Z - \frac{1}{2} Z_{\theta_1}^\prime \mathcal{I}_{\theta_1} Z_{\theta_1} - Z^\prime_{\rho_U \rho_U^\prime}  \mathcal{I}_{\rho_U \rho_U^\prime\theta_1} Z_{\theta_1} \\
& \quad - \frac{1}{2}Z^\prime_{\rho_U \rho_U^\prime} \mathcal{I}_{\rho_U \rho_U^\prime} Z_{\rho_U \rho_U^\prime} + Z^\prime_{\theta_1} \mathcal{I}_{\theta_1}  \hat{\tau}_{\theta_1,n} + Z^\prime_{\rho_U\rho_U^\prime} \mathcal{I}_{\rho_U\rho_U^\prime\theta_1}  \hat{\tau}_{\theta_1,n}-\frac{1}{2}\hat{\tau}^\prime_{\theta_1,n} \mathcal{I}_{\theta_1} \hat{\tau}_{\theta_1,n} \\
& \quad + Z^\prime_{\rho_U\rho_U^\prime,n} \mathcal{I}_{\rho_U\rho_U^\prime} \hat{\tau}_{\rho_U\rho_U^\prime,n} + Z^\prime_{\theta_1,n} \mathcal{I}_{\theta_1\rho_U\rho_U^\prime} \hat{\tau}_{\rho_U\rho_U^\prime,n} \\
& \quad - \hat{\tau}^\prime_{\rho_U\rho_U^\prime,n}  \mathcal{I}_{\rho_U\rho_U^\prime}  \hat{\tau}_{\theta_1,n} - \frac{1}{2} \hat{\tau}^\prime_{\rho_U\rho_U^\prime,n}  \mathcal{I}_{\rho_U\rho_U^\prime\theta_1} \hat{\tau}_{\rho_U\rho_U^\prime,n} + o_P(1)\\
=& \ell_n(\theta^\ast_0)  + \frac{1}{2} Z^\prime \mathcal{I} Z - \frac{1}{2} \left( \hat\tau_{\theta_1,n} - Z_{\theta_1}\right)^\prime \mathcal{I}_{\theta_1} \left( \hat\tau_{\theta_1,n} - Z_{\theta_1}\right) \\
&\quad  - \left(  \hat\tau_{\rho_U \rho_U^\prime,n} - Z_{\rho_U \rho_U^\prime}\right)^\prime \mathcal{I}_{\rho_U \rho_U^\prime \theta_1} \left( \hat\tau_{\theta_1,n} - Z_{\theta_1}\right) \\
& - \frac{1}{2} \left(  \hat\tau_{\rho_U \rho_U^\prime,n} - Z_{\rho_U \rho_U^\prime}\right)^\prime \mathcal{I}_{\rho_U \rho_U^\prime } \left(  \hat\tau_{\rho_U \rho_U^\prime,n} - Z_{\rho_U \rho_U^\prime}\right) \\
=& \ell_n(\theta^\ast_0)  + \frac{1}{2} Z^\prime \mathcal{I} Z - \frac{1}{2}\left(\hat\tau_n - Z \right)^\prime \mathcal{I} \left(\hat\tau_n - Z \right),
\end{align*}
where $\hat\tau_n = (\hat{\tau}^\prime_{\theta_1,n}, \hat{\tau}^\prime_{\rho_U \rho_U^\prime,n})^\prime$, and the statement of the Theorem follows from Assumption \ref{ass:sigmau0pos} and \citet[][Th. 4 p. 1365]{andrews1999}.

\section{Descriptive Statistics} \label{sec:appC}

Table \ref{tab:descrstat} contains descriptive statistics from the main variables used in the empirical analysis.
% latex table generated in R 4.0.4 by xtable 1.8-4 package
% Thu Mar 18 12:37:32 2021
\begin{table}[ht]
\centering
\begin{tabular}{ccccc}
  \hline
\hline
 & Mean & St.Dev. & Min & Max \\ 
  \hline
\hline
Output & 176516.050 & 295059.190 & 2466.286 & 3455000.000 \\ 
   \hline
\multicolumn{1}{l}{\textit{Inputs}} &  &  &  &  \\ 
   \hline
Land & 27266.243 & 31344.445 & 729.000 & 273800.000 \\ 
  Labor & 228.441 & 838.113 & 1.000 & 13374.000 \\ 
  Fertilizers & 41358.787 & 434911.849 & 50.000 & 7500000.000 \\ 
  Seeds & 277.489 & 392.705 & 0.004 & 3500.000 \\ 
   \hline
\multicolumn{1}{l}{\textit{Environmental variables}} &  &  &  &  \\ 
   \hline
Education & 0.065 & 0.138 & 0.000 & 0.800 \\ 
  Experience & 23.833 & 16.019 & 1.000 & 77.000 \\ 
  Risk Div & 0.571 & 0.575 & 0.002 & 3.051 \\ 
   \hline
\multicolumn{1}{l}{\textit{Instruments}} &  &  &  &  \\ 
   \hline
Natural Shock & 0.439 & 0.497 & 0.000 & 1.000 \\ 
  Own Supplier & 0.051 & 0.099 & 0.000 & 0.999 \\ 
  Formal Supplier & 0.256 & 0.194 & 0.000 & 1.000 \\ 
  Informal Supplier & 0.011 & 0.045 & 0.000 & 0.500 \\ 
  Peers Experience & 24.469 & 13.449 & 10.000 & 44.000 \\ 
   \hline
\hline
\end{tabular}
\caption{Descriptive Statistics} 
\label{tab:descrstat}
\end{table}

\end{document}